\documentclass[
 reprint,
 amsmath,amssymb,
 aps,
physrev,
 longbibliography
]{revtex4-2}

\usepackage[T1]{fontenc}
\usepackage{graphicx}
\usepackage{dcolumn}
\usepackage{bm}
\usepackage{xcolor}
\usepackage{booktabs}
\usepackage{hyperref}

\begin{document}

\title{Background-Limited Silicon Telescope Methodology for Weak Internal Pair Creation Transition in Low-Energy Deuteron-Deuteron Reactions}

\author{Gokul Das H$^1$, K. Czerski$^1$, S. Thulichery$^1$, R. Dubey$^1$, M. Valat$^1$, N. Targosz-\'Sl{\k e}czka$^1$, A. Kowalska$^2$ }
\affiliation{$^1$Institute of Physics, University of Szczecin, 70-451 Szczecin, Poland \\ $^2$Physics Department, Maritime University of Szczecin, 70-500 Szczecin, Poland}


\begin{abstract}
A background-constrained silicon-telescope methodology has been developed  to investigate recently observed internal $e^-e^+$ pair creation in low-energy deuteron-deuteron fusion. The decay of the near-threshold $0^+$ resonance in $^4\mathrm{He}$ produced in DD fusion results in an $e^-e^+$ pair sharing a total energy of $22.84$ MeV. These high-energy leptons traverse thin silicon detectors and deposit only part of their kinetic energy. The resulting partial-absorption peak structure overlaps with the natural-radiation background of cosmic-ray secondaries and terrestrial gamma rays. The silicon detector telescope used in the present experiments is characterised against this background through a component-by-component model that combines a GEANT4 cosmic-ray simulation, a large-volume NaI(Tl)-constrained terrestrial-gamma propagation, and a beam-correlated neutron response. Long-duration measurements in multiple orientations and shielding configurations are used to quantify the run-to-run variability of the background and to translate it into an explicit systematic envelope on the extracted signal. The framework is then applied to the beam-on singles spectrum and the two-dimensional coincidence response, yielding a quantitative branching-ratio measurement under background-limited conditions. The methodology is directly transferable to other weak-channel searches in surface laboratories where continuum-like signals overlap with natural-radiation backgrounds.
\end{abstract}

\maketitle

\section{Introduction}

Deuteron--deuteron (DD) fusion is important for both prospective fusion-energy concepts and energy generation in dense astrophysical environments, including giant planets, brown dwarfs, and white-dwarf supernova progenitors \cite{ouyed2016,ichimaru1999}. Far below the Coulomb barrier, the bare DD reaction cross section decreases exponentially with decreasing projectile energy. Laboratory studies with gaseous and metallic targets have shown that surrounding electrons can partially screen the nuclear charges and enhance the reaction cross section by reducing the effective Coulomb barrier. This electron-screening effect is especially strong in metallic deuterides, where quasifree conduction electrons and lattice effects can produce screening energies much larger than those observed in gaseous targets \cite{Czerski2001,Kasagi2002,agata2025}. Measurements of low-energy DD fusion in metallic targets are therefore relevant both for understanding nuclear reactions in screened laboratory environments and for testing how screening may modify reaction rates in dense astrophysical conditions.

Recent studies have suggested that the low-energy enhancement of the DD fusion reaction cannot be explained by electron screening alone. Measurements of the $^{2}\mathrm{H}(d,p)^{3}\mathrm{H}$ reaction at deuteron energies below $10~\mathrm{keV}$ indicate that an additional contribution from a narrow $J^{\pi}=0^{+}$ threshold resonance in $^{4}\mathrm{He}$ is required to account for the observed enhancement \cite{czerski2022prc}. This state is interpreted as a single-particle resonance with a $2+2$ cluster structure, arising from the weak coupling between the $1+3$ and $2+2$ cluster configurations in the compound $^{4}\mathrm{He}$ nucleus. Because of its single-particle character, the resonance has a strongly energy-dependent entrance-channel width governed by the Coulomb penetration factor, and can significantly alter the branching ratios of the DD reaction at very low energies. In particular, theoretical estimates based on the $E_0$ energy-weighted sum rule predict that the resonance should have a large partial width for the decay to the $^{4}\mathrm{He}$ ground state through internal electron--positron pair creation (IPC), producing a continuous energy spectrum extending up to approximately $23~\mathrm{MeV}$ \cite{czerski2022prc}. The contribution of this internal-pair-creation channel to the total DD reaction cross section is expected to increase with decreasing deuteron energy far below the Coulomb barrier, eventually dominating over the proton and neutron channels at energies below approximately $5~\mathrm{keV}$.

Experimental indications of high-energy electron and positron emission have been reported using single silicon detectors of different thicknesses, supported by GEANT4 simulations of electron transport and partial energy deposition \cite{czerski2024prc,gokul2024,prx2025}. In these measurements, the high-energy leptons are not fully stopped in the silicon detector. Instead, they traverse the active volume and deposit only part of their kinetic energy, producing a broad partial-absorption structure whose position scales with the detector thickness, located at approximately $84~\mathrm{keV}$ for a $300~\mu\mathrm{m}$ detector and at $980~\mathrm{keV}$ for a $3~\mathrm{mm}$ detector \cite{gokul2024}. Since charged particles continuously lose energy while passing through matter, each electron or positron crossing the active silicon volume produces a detectable signal, giving an intrinsic $e^{-}e^{+}$ detection efficiency close to unity. This makes compact silicon detectors well suited for measuring the IPC channel with high efficiency. An additional advantage is that the same detector records both the IPC events and the charged DD reaction products, such as protons, allowing the electron/proton branching ratio to be determined directly from the measured count ratio  with only relative-efficiency corrections. More recent measurements combining silicon detectors with large-volume NaI(Tl) and HPGe detectors have further linked the charged-particle spectra to bremsstrahlung and annihilation signatures from energetic positrons \cite{prx2025}.

Despite this progress, extending the electron/proton branching-ratio measurement to deuteron energies below $10~\mathrm{keV}$ is limited by two factors. First, the DD fusion cross section decreases exponentially with decreasing energy, so that the absolute event yield remains small even at high beam currents and long counting times. Second, the electron and positron response in silicon detector lies in an energy region where natural radiation backgrounds are significant, and the IPC signal itself is broad and continuum-like, preventing isolation through a simple region-of-interest analysis. In previous measurements, the branching ratio was extracted by subtracting experimentally acquired background spectra from beam-on data. While effective at higher energies where the signal is clearly visible above background, this approach becomes fragile when the yield drops, because small changes in the background rate or spectral shape can bias the extracted IPC contribution. A quantitative, model-based understanding of the silicon-detector background is therefore required to push the branching-ratio measurement below the previously accessible energy range.

The dominant background sources in surface-laboratory silicon detectors are cosmic-ray secondaries and terrestrial gamma radiation. Cosmic-ray muons, electrons, and other charged secondaries traverse the silicon and produce partial energy depositions similar to the IPC response. Terrestrial gamma rays from the decay of $^{40}\mathrm{K}$, $^{232}\mathrm{Th}$, and $^{238}\mathrm{U}$ and natural alpha-decay series in building materials and surrounding structures interact mainly through Compton scattering and generate a broad continuum increasing sharply toward lower energies. Both background components can vary over time. The cosmic-ray flux varies with atmospheric conditions \cite{Muonvariation, MuonSeasonalMINOS}, and departures from secular equilibrium through radon escape can introduce time-dependent changes to the gamma-ray intensity \cite{mauring2014,markovic2019}. Low-background techniques addressing these sources have been developed extensively for HPGe gamma spectrometry and rare-event searches. For silicon detectors specifically, earlier work established the roles of cosmic rays, gamma sensitivity, and material contamination \cite{walter1966}, while more recent studies have addressed muon-induced backgrounds in low-background alpha spectroscopy \cite{misiaszek2013}, terrestrial-radiation-dominated backgrounds in airborne monitoring systems \cite{becker2021,grujic2011}, and the full background spectrum in silicon substrates for superconducting quantum circuits \cite{fowler2024prxq}.

In most of these applications, the particles of interest deposit their full energy within the detector. High-energy backgrounds such as cosmic-ray muons, which traverse the detector without stopping, can therefore be rejected using active veto techniques \cite{misiaszek2013}. The methodology proposed in this paper differs in a fundamental way. The IPC electrons and positrons also traverse the thin silicon detectors without stopping, depositing only a fraction of their kinetic energy, and produce partial-absorption signatures similar to those of penetrating cosmic-ray secondaries. An active veto that rejects all traversing particles would therefore suppress the signal itself. A coincidence requirement between the two detectors of the telescope can, however, suppress the terrestrial gamma-ray component, which interacts predominantly in a single detector through Compton scattering. The background characterization problem is therefore to quantitatively model the spectral shape of the background response in the energy region where the IPC signal is expected, so that the two contributions can be reliably separated.

The present work addresses these challenges by employing a $\Delta E$--$E$ silicon detector telescope, consisting of a $500~\mu\mathrm{m}$ and a $1~\mathrm{mm}$ thick PIPS detector in close geometry, to measure the DD reaction products at very low energies. The telescope geometry provides two complementary modes of analysis. In the singles mode, the $1~\mathrm{mm}$ detector spectrum is used to extract the electron/proton branching ratio through a model-based fit of the individually simulated background and signal response functions. In the coincidence mode, events recorded simultaneously in both detectors provide an independent observable with substantially different background composition.

The cosmic-ray and terrestrial gamma-ray backgrounds are simulated independently and validated against long-duration measurements in multiple detector configurations, allowing the stability and relative composition of the background to be assessed for both the singles and coincidence observables.

The paper is organized as follows. Section~\ref{sec:setup} describes the experimental setup, detector telescope, accelerator configuration, and data acquisition. Section~\ref{sec:detector} presents the detector optimization, calibration, and GEANT4 validation. Section~\ref{sec:background} gives the cosmic-ray, terrestrial-gamma, and long-term background characterization. Section~\ref{sec:signal} describes the signal-extraction framework and its application to the DD beam data. Section~\ref{sec:discussion} discusses the implications for future weak-channel measurements in low-energy DD fusion and for silicon-detector background studies in surface laboratories.

\section{Experimental Setup}
\label{sec:setup}

\textit{Beam measurements:} The DD fusion measurements were performed at the eLBRUS ultra-high-vacuum accelerator facility at the University of Szczecin~\cite{mateuzaccelerator,czerski2024prc}. The accelerator is equipped with an electron cyclotron resonance ion source and a magnetic analyzing system, which allows different deuterium ion species, $\mathrm{D}^{+}$, $\mathrm{D}_{2}^{+}$, and $\mathrm{D}_{3}^{+}$, to be separated and selected before transport to the target chamber. In the present experiment, monoatomic $\mathrm{D}^{+}$ beams with laboratory energies between $10$ and $20~\mathrm{keV}$ were delivered to the target. The beam current was typically $45~\mu\mathrm{A}$ for the $18~\mathrm{keV}$ run used in the singles-spectrum branching-ratio analysis, while the average current over the $10$--$20~\mathrm{keV}$ data set used for the two-dimensional coincidence analysis was approximately $53~\mu\mathrm{A}$. The beam was incident on a $0.2 ~\mathrm{mm}$-thick palladium target mounted at $45^{\circ}$ with respect to the beam direction, after passing through a circular collimator of approximately $ 5~\mathrm{mm}$ radius. The target chamber was maintained under ultra-high-vacuum conditions, typically at $10^{-9}~\mathrm{mbar}$ with a beam on the target, to minimize surface contamination during long irradiations.

Charged particles emitted from the target were detected using a compact silicon detector telescope operated in a $\Delta E$--$E$ configuration. The telescope consisted of two Passivated Implanted Planar Silicon (PIPS) detectors manufactured by Mirion Technologies. The front detector, referred to as the $\Delta E$ detector, was a fully depleted $500~\mu\mathrm{m}$-thick PIPS detector with an active area of $300~\mathrm{mm}^{2}$. The rear detector, referred to as the $E$ detector, was a partially depleted $1~\mathrm{mm}$-thick PIPS detector with the same active area. The two detectors were mounted with a separation of approximately $2~\mathrm{mm}$, and the telescope was positioned at a distance of approximately $5~\mathrm{cm}$ from the target. The ordinary charged DD reaction products, such as protons, tritons, and $^{3}\mathrm{He}$ ions, are stopped in the $\Delta E$ detector. In contrast, high-energy electrons and positrons from the IPC channel pass through both detectors, depositing only a fraction of their energy in each. The high-energy $e^{-}e^{+}$ events can therefore be identified through coincidences between the $\Delta E$ and $E$ detectors. This allowed the telescope to record both the ordinary charged-particle channels and the expected IPC channel response in the same measurement geometry. A thin aluminum absorber foil of $0.8~\mu\mathrm{m}$ thickness was placed in front of the telescope. The foil was used to suppress backscattered deuterons reaching the silicon detectors, while remaining thin enough to transmit the charged DD reaction products and high-energy electrons with minimal energy loss.

The detector signals were processed using a CAEN DT5780P digitizer with 14-bit resolution and a sampling rate of $100~\mathrm{MHz}$. Data acquisition was performed with the CoMPASS software package. Events were recorded in list mode, including both pulse-height and timestamp information for each detector channel. The data were stored in ROOT format and analyzed offline. Singles spectra were constructed independently for the $\Delta E$ and $E$ detectors, with the $E$ detector spectrum serving as the primary observable for the branching ratio determination. Coincidence events were identified offline using the timestamp difference between the two detector signals. The timing selection used to define true $\Delta E$--$E$ coincidences is described in Sec.~III.

\textit{Background measurements:} In addition to the beam measurements, dedicated measurements were performed to characterize the natural-radiation background seen by the silicon detectors. These measurements were first carried out in a preparation room located approximately $2~\mathrm{m}$ from the accelerator hall, using the same $\Delta E$--$E$ telescope and readout system as in the beam experiment. The telescope was mounted inside a small cylindrical stainless-steel vacuum chamber with an outer diameter of $11.5~\mathrm{cm}$, a height of $17~\mathrm{cm}$, and a wall thickness of $0.7~\mathrm{cm}$, evacuated to approximately $10^{-3}~\mathrm{mbar}$ using an oil-free diaphragm pump. Background spectra were recorded with the telescope in horizontal and vertical orientations. For shielded measurements, the chamber containing the telescope was placed inside a passive low-background enclosure constructed from low-activity lead with an $11~\mathrm{cm}$ wall thickness, internally lined with $0.5~\mathrm{cm}$ of copper and $0.5~\mathrm{cm}$ of plexiglass to suppress secondary fluorescence from the lead. The horizontal and vertical orientation measurements were repeated inside this shielding.

The ambient gamma-ray field in the preparation room was characterized independently using a large-volume $8'' \times 12''$ NaI(Tl) scintillation detector read out by four photomultiplier tubes. The measured gamma spectrum was used to constrain the terrestrial component of the silicon detector background model, as described in Sec.~V.

Long-duration background measurements were also performed inside the accelerator chamber with the telescope installed in the beam-experiment geometry. As an independent check of possible instrumental effects on the observed background variability, the accelerator-chamber measurement was repeated with a separate $1~\mathrm{mm}$-thick PIPS detector with an active area of $150~\mathrm{mm}^{2}$, read out through an independent analog electronics chain and a TUKAN 8K multichannel analyzer. The analysis, modelling, and interpretation of all background measurements are presented in Secs.~IV and V.

\section{Detector Response and Efficiency Calibration}
\label{sec:detector}

The electron/proton branching-ratio extraction requires a reliable detector response function for high-energy IPC $e^{-}e^{+}$ pairs. Because these leptons are not fully stopped in the Si detectors, the measured response depends on partial energy deposition, detector thickness, thresholds, coincidence conditions, and scattering in nearby materials. These effects were modeled with GEANT4 Monte Carlo simulations. This section describes the simulation framework, detector-thickness optimization, energy and coincidence definitions, validation with calibration sources, and the simulated IPC efficiency factors used in the signal extraction.

\subsection{GEANT4 simulation framework}
\label{sec:geant4_framework}

The detector response was simulated using the GEANT4 Monte Carlo particle transport toolkit.\cite{GEANT4} The geometry was adapted to each measurement configuration. For calibration measurements performed in the preparation room, the model included the two PIPS detectors with their housings and packaging, the radioactive source geometry as specified by the manufacturer, and the cylindrical stainless-steel vacuum chamber described in Sec.~II. For shielded configurations, the lead enclosure with its copper and plexiglass lining was additionally included. For measurements in the accelerator environment, the geometry comprised the palladium target, molybdenum target holder, aluminum absorber foil, the ultra-high-vacuum chamber with its $0.5~\mathrm{cm}$ stainless-steel walls and $0.2~\mathrm{cm}$ mu-metal lining, and the full detector telescope. The physics models, transport settings, and analysis procedures were common to all configurations and only the geometry was changed to match each measurement.

A custom modular physics list was used in the simulations. Electromagnetic interactions were modelled using the G4EmStandardPhysics\_option4 constructor, which provides the most accurate low-energy electromagnetic transport available in GEANT4 \cite{Ivanchenko}. Atomic de-excitation was enabled to include X-ray fluorescence, Auger electrons, and Auger cascades in detector materials and surrounding structures. Hadronic interactions were modelled using FTFP\_BERT\_HP, which was required for the neutron response calculations used in the beam-on analysis. Production cuts of $1~\mu\mathrm{m}$ were applied to electrons, positrons, protons, and photons. For each simulated event, the energy deposited in the active volumes of the $\Delta E$ and $E$ detectors was recorded separately, and simulated singles spectra and two-dimensional $\Delta E$--$E$ histograms were constructed using the same thresholds and selection conditions as the experimental data.

\subsection{Detector-thickness optimization}
\label{sec:thickness_optimization}

The IPC electrons and positrons, with energies extending up to approximately $23~\mathrm{MeV}$, pass through the silicon detectors without stopping. The observed signal is therefore a partial-absorption structure whose position depends on the detector thickness \cite{gokul2024}. If the detector is too thin, this structure falls close to or below the electronic noise threshold.

GEANT4 simulations were performed for front detector thicknesses of $100$, $300$, and $500~\mu\mathrm{m}$, each paired with a $1~\mathrm{mm}$ rear detector. A lower threshold of approximately $100~\mathrm{keV}$ was adopted as the design criterion, based on the stable low-noise operating conditions achieved in the preparation room measurements. The simulations showed that $100$ and $300~\mu\mathrm{m}$ front detectors place the IPC partial-absorption peak close to or below this threshold, resulting in severe coincidence sensitivity losses. The $500~\mu\mathrm{m}$ front detector shifts the partial-absorption response above the threshold, while the $1~\mathrm{mm}$ rear detector places its corresponding response near $300$--$400~\mathrm{keV}$, well above the noise threshold and in a spectral region with relatively low background contamination. The final telescope configuration of $500~\mu\mathrm{m}$ $(\Delta E)$ + $1~\mathrm{mm}$ $(E)$ was selected on this basis.

\begin{figure}[t]
\centering
\includegraphics[width=0.95\columnwidth]{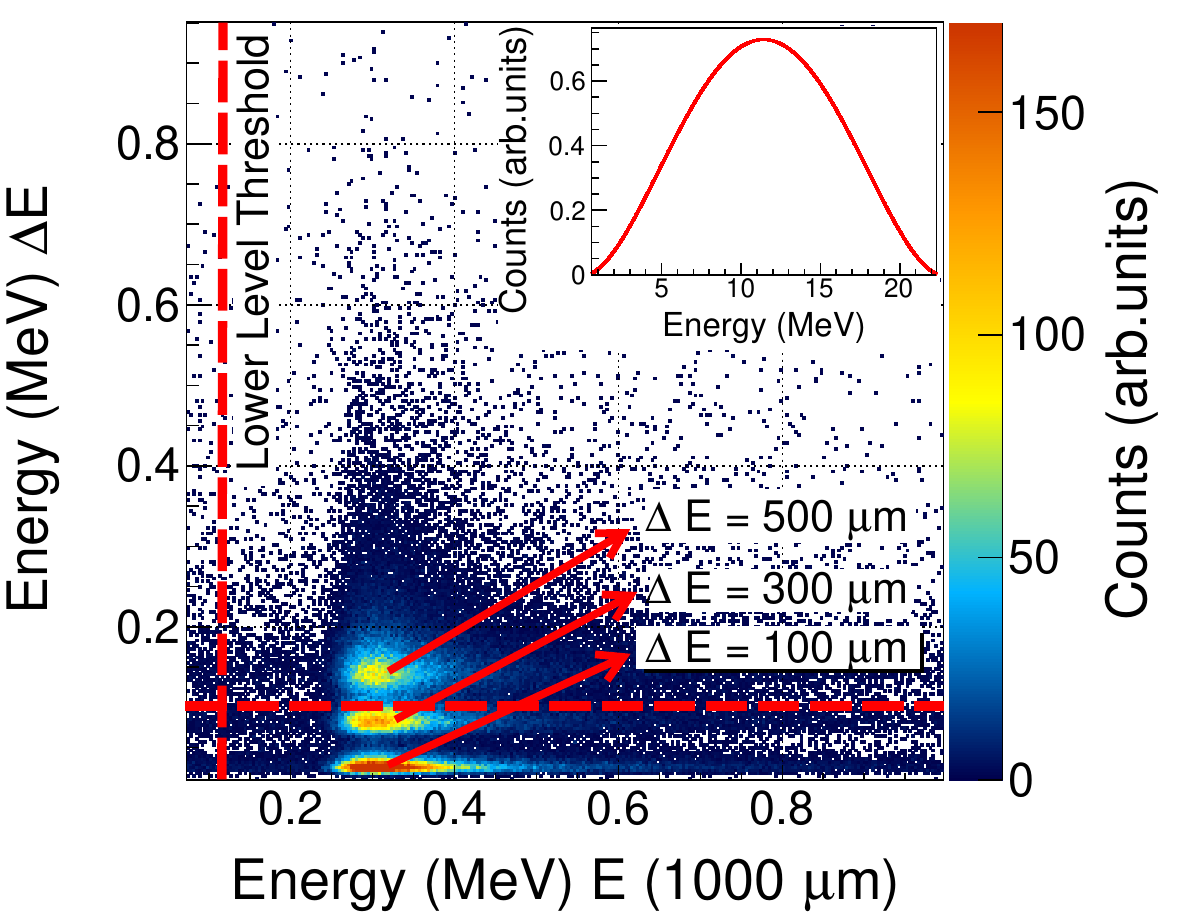}
\caption{GEANT4 simulated $\Delta E$--$E$ spectrum for IPC electrons. The dashed lines indicate the lower-level threshold used in the detector optimization. The simulated response is shown for different $\Delta E$-detector thicknesses, demonstrating that the $500~\mu\mathrm{m}$ detector places the IPC partial-energy-deposition response above threshold.}
\label{fig:ipc_thickness_optimization}
\end{figure}

\subsection{Energy calibration and coincidence definition}
\label{sec:energy_calibration}

The energy calibration of both detectors was established using a $^{239}\mathrm{Pu}$--$^{241}\mathrm{Am}$--$^{244}\mathrm{Cm}$ triple-alpha source, with the alpha lines at $5.157$, $5.486$, and $5.805~\mathrm{MeV}$ providing reference points. The calibration in the electron energy region was extended using $^{60}\mathrm{Co}$ and $^{204}\mathrm{Tl}$ sources. During beam operation, the $\Delta E$ detector calibration was verified using the proton, triton, and $^{3}\mathrm{He}$ lines from the DD reaction, confirming that the calibration parameters remained valid under accelerator conditions.

\begin{figure}[t]
\centering
\includegraphics[width=0.99\columnwidth]{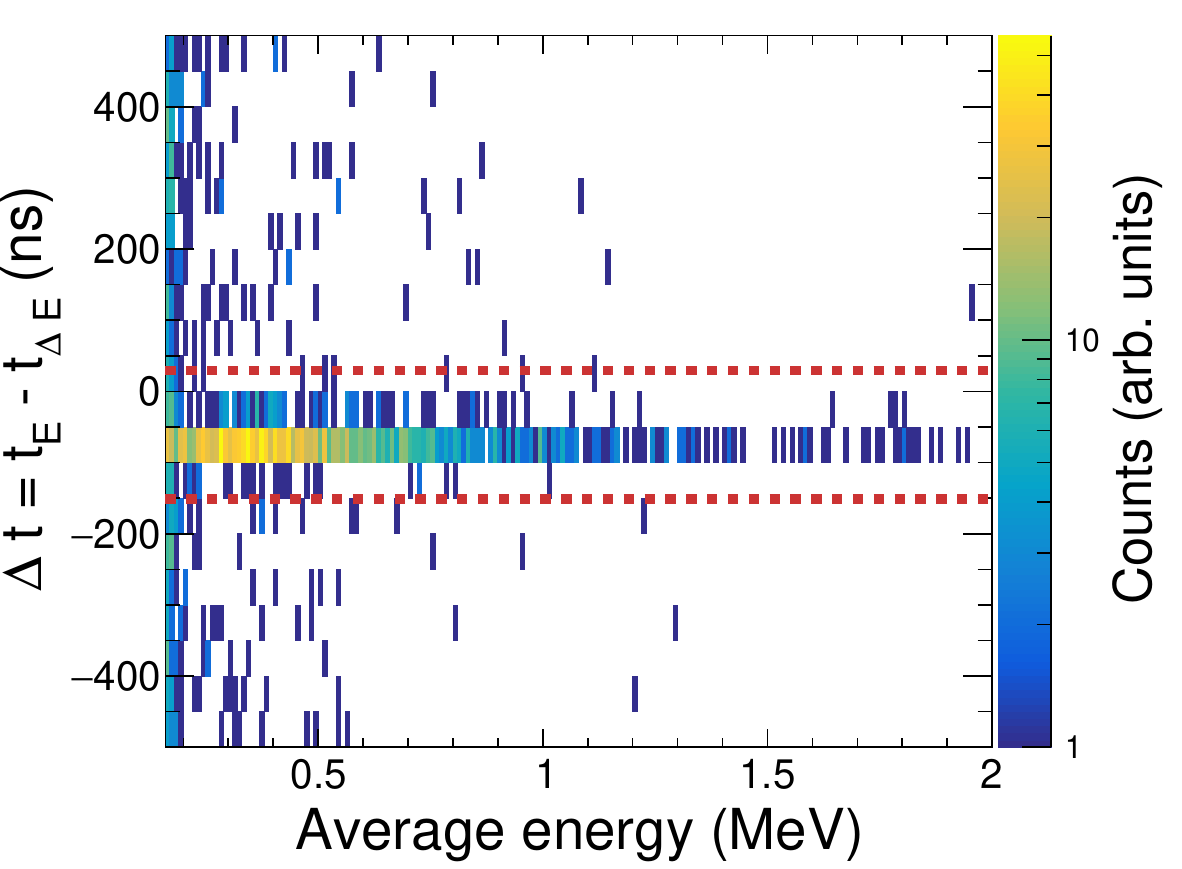}
\caption{Time-difference spectrum between the $E$ and $\Delta E$ detector signals as a function of the average deposited energy. The dashed lines indicate the prompt coincidence window used to define true $\Delta E$--$E$ coincidences.}
\label{fig:ctof}
\end{figure}

Coincidence events were defined offline using the timestamp information recorded for each detector channel. For each candidate event pair, the time difference
\[
\Delta t = t_{E} - t_{\Delta E}
\]
was calculated and plotted as a function of the average deposited energy
\[
E = \frac{E_{\Delta E} + E_{E}}{2}.
\]
This two-dimensional representation reveals a well-defined prompt coincidence band, clearly separated from the uncorrelated background, as shown in Fig.\ref{fig:ctof}. 

The accepted time window was determined from the boundaries of this prompt band and applied consistently to calibration data, background measurements, and beam-on data. All two-dimensional $\Delta E$--$E$ histograms presented in the following sections were constructed using this same coincidence condition.

\subsection{Validation of simulation and IPC efficiency determination}
\label{sec:simulation_validation}

\begin{figure*}[t]
\centering
\includegraphics[width=0.75\textwidth]{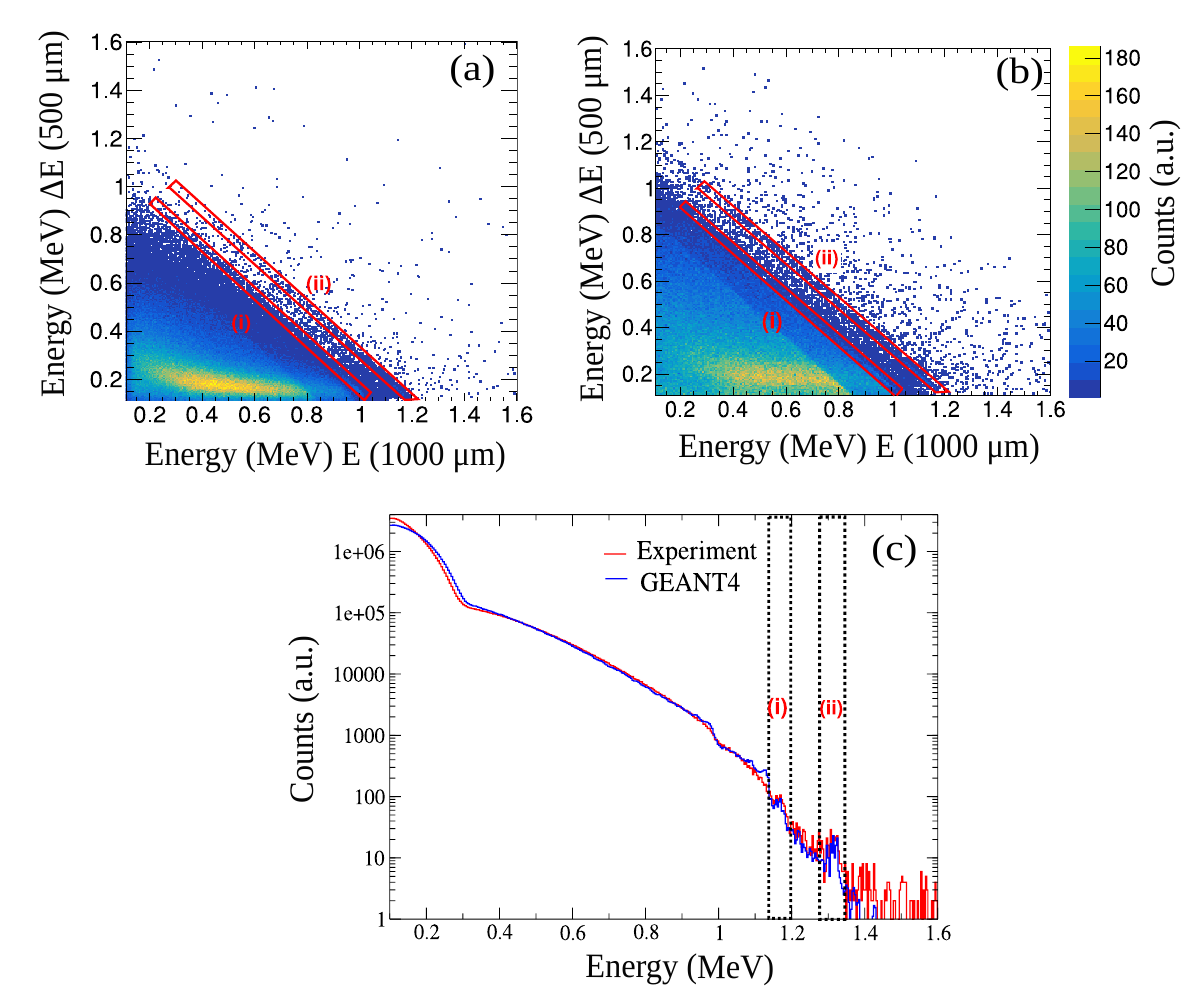}
\caption{Comparison between experimental and simulated $^{60}\mathrm{Co}$ electron response in the silicon telescope. Panels (a) and (b) show the $\Delta E$--$E$ coincidence spectra for experiment and GEANT4 simulation, respectively. Panel (c) compares the corresponding one-dimensional spectra used for the validation of the simulated electron transport and partial-energy-deposition response.}
\label{fig:co60_validation}
\end{figure*}

The GEANT4 simulatons of electron transport and partial energy deposition in the telescope was validated experimentally using a $^{60}\mathrm{Co}$ beta  and internal conversion source. A direct validation using an IPC electron source is not feasible, since no calibrated source producing a continuous high-energy electron-positron spectrum analogous to the IPC decay is practically available. The $^{60}\mathrm{Co}$ source provides the closest practical alternative as its internal conversion electrons at $1.162$ and $1.320~\mathrm{MeV}$ are monoenergetic and sufficiently energetic to penetrate the $\Delta E$ detector, depositing only a fraction of their energy as a partial-absorption signal, while being fully stopped in the $1~\mathrm{mm}$ $E$ detector. This produces clean, identifiable structures in both the singles and coincidence spectra that can be compared directly between experiment and simulation as shown in Fig.~\ref{fig:co60_validation}.

To compare experiment and simulation quantitatively, the validation ratio
\[
f=\frac{N_{2D} }{N_{1D}}
\]
was constructed for each conversion electron line, where $N_{\mathrm{1D}}$ is the number of counts in the full-energy peak in the $\Delta E$ singles spectrum and $N_{\mathrm{2D}}$ is the number of counts in the corresponding diagonal gate in the $\Delta E$--$E$ histogram. Identical gate boundaries were applied to both the experimental and simulated spectra. This ratio is independent of the absolute source activity and source-detector distance, since both quantities are measured simultaneously in the same geometry, and it depends only on the detector thicknesses and relative geometry. The measured and simulated values are summarized in Table~ \ref{tab:co60_validation}.

\begin{table}[t]
\caption{Comparison between experimental and simulated validation ratios for the $^{60}\mathrm{Co}$ conversion-electron lines.}
\label{tab:co60_validation}
\begin{ruledtabular}
\begin{tabular}{ccc}
Electron energy (MeV) & $f$ Experiment (\%) & $f$ Simulation (\%) \\
\hline
$1.17$ & $8.4 \pm 0.4$ & $8.9 \pm 0.2$ \\
$1.33$  & $10.6 \pm 0.9$ & $11.0 \pm 0.6$ \\
\end{tabular}
\end{ruledtabular}
\end{table}

Using the validated simulation, the detection efficiency of the telescope for IPC electrons and positrons was calculated for the beam-experiment geometry. The continuous IPC electron-positron energy spectrum extending up to $22.84~\mathrm{MeV}$ [Fig.~1, inset] was used as input, with $10^{8}$ pairs generated isotropically from the target surface and transported through the full accelerator chamber geometry. The same energy thresholds and analysis conditions applied to the experimental data were imposed on the simulated events.

The quantity required for the branching ratio determination is the conditional geometric efficiency $f(dE \mid E)$, defined as the probability that an IPC electron producing a signal in the $\Delta E$ detector also produces a signal in the $E$ detector. This quantity accounts for the small difference in solid angle subtended by the two detectors, which are separated by $2~\mathrm{mm}$. Since both detectors have near- 100\% intrinsic detection efficiency for both protons and electrons \cite{gokul2024}, $f(dE \mid E)$ is the only efficiency correction entering the branching ratio calculation. The simulated value is $f(dE\mid E)=0.95\pm0.06,$
 and its systematic uncertainty is propagated from the $^{60}\mathrm{Co}$ validation result. It is applied in the branching ratio extraction described in Sec.~V.
 \section{Background Characterization}
\label{sec:background}

The branching ratio extraction requires a quantitative understanding of the background response in the energy region where the IPC signal is expected. In a surface laboratory, two natural sources of ionizing radiation contribute a continuous background to silicon detector spectra: secondary cosmic radiation reaching ground level, and terrestrial gamma radiation from the decay of primordial radioactive isotopes in surrounding materials.

Cosmic-ray secondaries arise from interactions of primary protons and heavier nuclei with nuclei in the upper atmosphere, producing extensive air showers that propagate to ground level. The ground-level flux is composed of muons, neutrons, protons, electrons, positrons, and photons ~\cite{Grieder2001,PDG2024,Sato2015}. Of these, muons dominate the silicon detector response because of their high penetrating power and low stopping power ~\cite{PDG2024,Bogdanova2006}. At typical ground-level energies (a few GeV), they traverse the full thickness of the silicon detectors without stopping, depositing only a fraction of their kinetic energy  ~\cite{Groom2001}. The other shower components contribute at lower levels through direct ionization, secondary production in surrounding materials, and electromagnetic cascades~\cite{Grieder2001}.

Terrestrial gamma radiation originates from the decay chains of the primordial isotopes $^{40}\mathrm{K}$, $^{232}\mathrm{Th}$, and $^{238}\mathrm{U}$, which are present in building materials, underlying soil, and rock formations. These chains produce gamma-ray emissions with energies extending up to $2.615~\mathrm{MeV}$. In thin silicon detectors, these photons interact predominantly through single Compton scattering, depositing only a fraction of their energy before the scattered photon escapes the active volume. The resulting energy spectrum is a broad continuum that overlaps with the expected IPC electron signal region.

Dedicated background measurements were performed with the $\Delta E$--$E$ telescope in multiple configurations - horizontal and vertical orientations, with and without lead shielding, in the preparation room and inside the accelerator chamber. Both cosmic and terrestrial background components were modeled independently and validated against these measurements. Because their contributions are present simultaneously in every measured spectrum and may exhibit different temporal behavior, the validation is treated jointly in Sec.~IV C, where the combined model is compared with the measured spectra and the temporal stability of each component is characterized.
\subsection{Cosmic-ray background model}
\label{sec:cosmic_background_model}

The cosmic-ray contribution was simulated using GEANT4 coupled to the Cosmic-ray Shower Library (CRY) version 1.7 as the primary particle source~\cite{CRY}. CRY generates realistic secondary cosmic-ray particle distributions at ground level, returning the particle species, kinetic energy, position, momentum direction, and arrival time for each secondary in the shower. Because the laboratory is located at approximately $20~\mathrm{m}$ above sea level, the sea-level altitude option in CRY was selected. The CRY configuration specified the geographic latitude of $53.44^{\circ}$ and the date of the experimental campaign to account for geomagnetic cutoff and solar-cycle modulation of the primary cosmic-ray flux.

\begin{figure}[t]
\centering
\includegraphics[width=0.95\columnwidth]{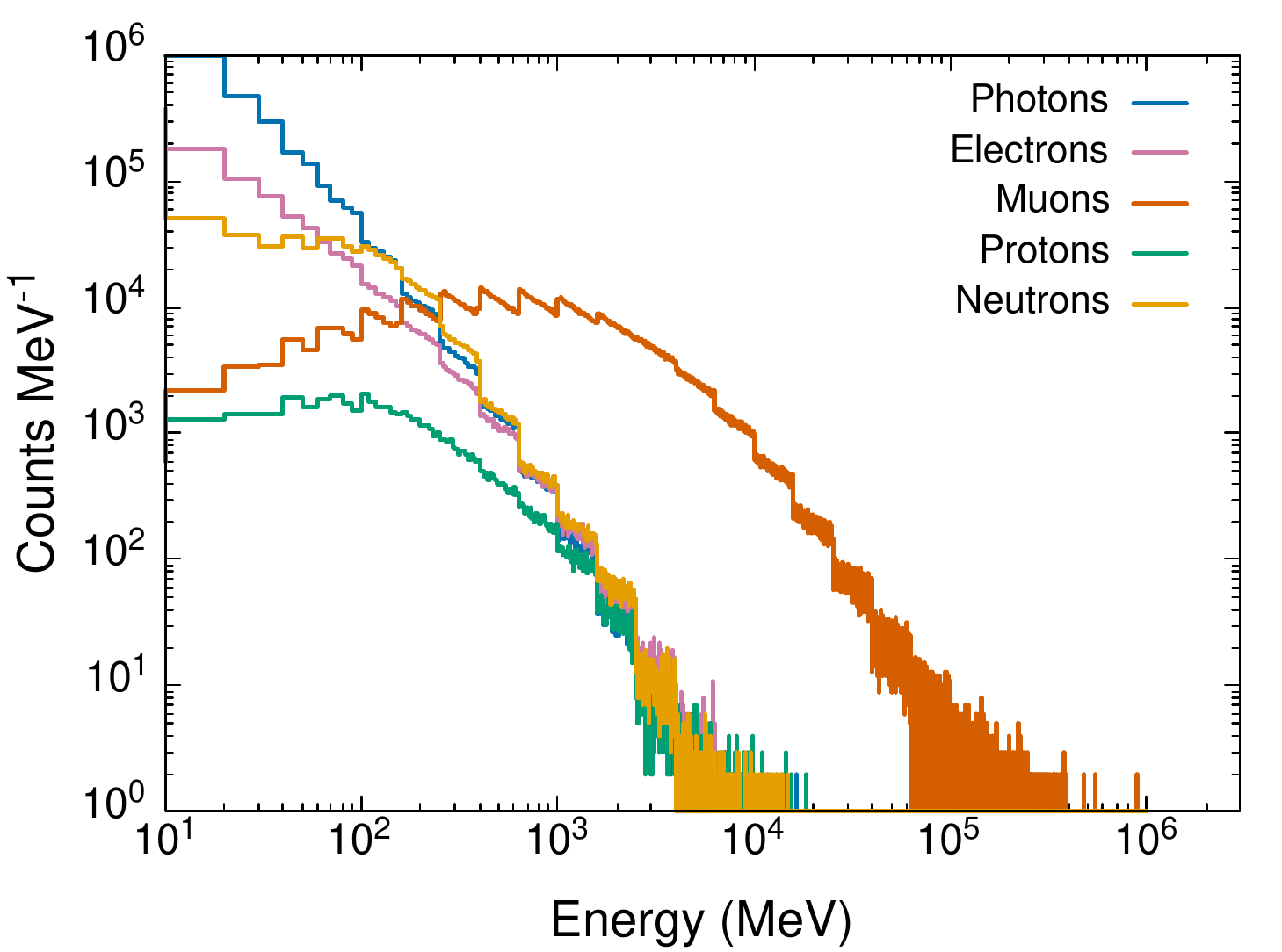}
\caption{Energy spectra of the cosmic-ray secondaries generated by CRY and used as input for the GEANT4 simulations.}
\label{fig:cry_input_spectra}
\end{figure}

The energy spectra of the cosmic-ray secondaries generated by CRY, used as input for the simulations are shown in Fig.~\ref{fig:cry_input_spectra}. Muons exhibit a hard spectrum extending into the GeV range with a mean energy of approximately $4~\mathrm{GeV}$, accounting for their high penetrating power. The electron and photon components are softer and concentrated below $100~\mathrm{MeV}$. These spectra serve as the input to the GEANT4 transport simulation.

Showers were generated on a $1~\mathrm{m} \times 1~\mathrm{m}$ horizontal plane positioned $2~\mathrm{mm}$ above the simulated concrete ceiling. For each GEANT4 event, CRY generated one complete shower, and all particles in that shower were transported through the detector geometry. The cumulative simulated live time reported by CRY was recorded and used to normalize the resulting spectra to the experimental acquisition time.

The GEANT4 geometry was adapted to each measurement configuration as described in Sec.~\ref{sec:geant4_framework}. Each configuration was simulated in the corresponding experimental orientation: horizontal and vertical in the preparation room, and $45^{\circ}$ with respect to the vertical in the accelerator chamber. A total of $10^{8}$ shower events were simulated per configuration. To decompose the detector response by particle species, an ancestry-based tracking scheme was implemented. When a cosmic-ray secondary entered a detector volume, its particle identity was recorded and propagated to all subsequent particles produced within that volume, allowing energy deposits to be attributed to the species of the entering particle.

\begin{figure*}[t]
\centering
\includegraphics[width=0.95\textwidth]{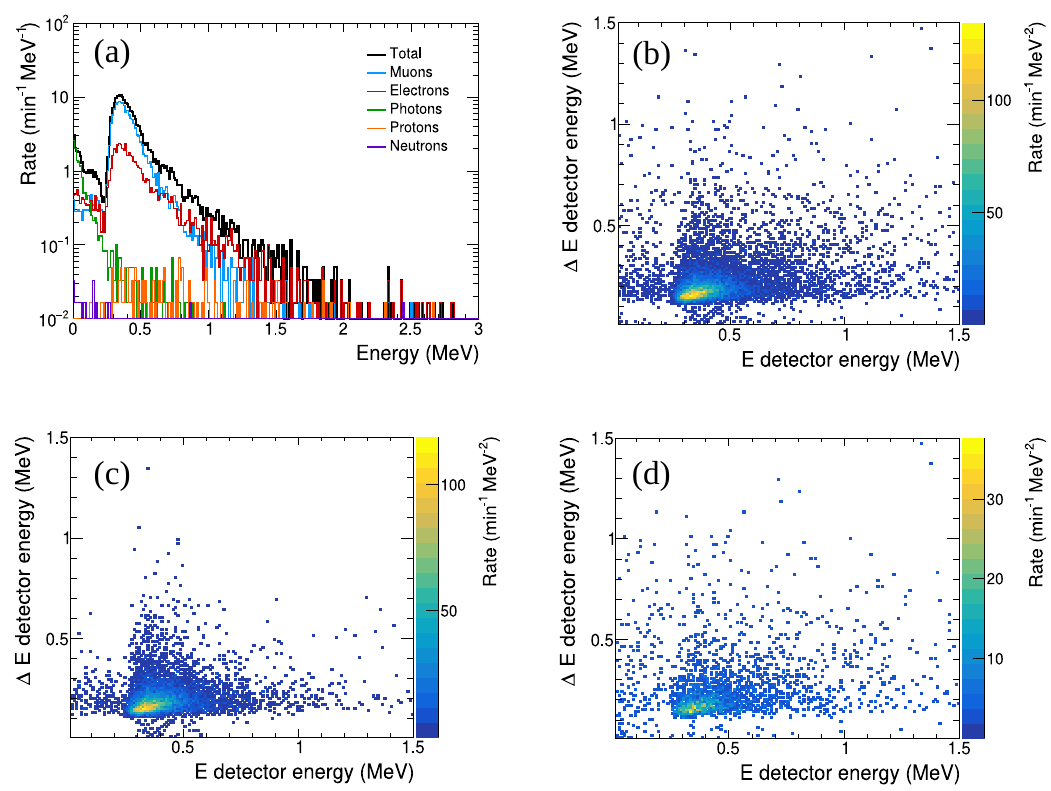}
\caption{Simulated cosmic-ray response in the horizontal preparation room configuration. Panel (a) shows the singles spectrum recorded in the $1~\mathrm{mm}$ $E$ detector, with the species decomposition obtained from the ancestry-based tracking scheme. Panel (b) shows the corresponding two-dimensional $\Delta E$--$E$ coincidence histogram. Panels (c) and (d) decompose this 2D response into the muon and electron contributions respectively.}
\label{fig:cosmic_response_4panel}
\end{figure*}

Figure~\ref{fig:cosmic_response_4panel} presents the simulated cosmic-ray response in the horizontal preparation room configuration. Panel (a) shows the singles spectrum in the $1~\mathrm{mm}$ $E$ detector, with the species decomposition obtained from the ancestry-based tracking scheme. The response forms a broad partial-absorption peak centered near $350~\mathrm{keV}$. The peak region is dominated by muons, whose contribution falls rapidly toward higher energies, while the high-energy cosmic-ray electron component produces a characteristic long tail that dominates over the muon contribution above $700~\mathrm{keV}$. Photons contribute primarily at low deposited energies, and protons and neutrons contribute only marginally to the total response.

Panel (b) shows the corresponding two-dimensional $\Delta E$--$E$ coincidence response. The coincidence requirement strongly suppresses the photon component, since gamma rays interact predominantly through isolated Compton events in a single detector and rarely produce correlated signals in both. The remaining 2D response is therefore produced mainly by muons and electrons that pass through both silicon detectors. Panels (c) and (d) show the muon and electron components separately. The muon component defines the dominant correlated band in the $\Delta E$--$E$ plane, while the electron component contributes a smaller but still visible diffuse population. The other particle species make only minor contributions to the coincidence response and are not shown separately.

The structure of the 2D response is important for the branching ratio analysis. The cosmic-ray distribution occupies a region in the $\Delta E$--$E$ plane that partially overlaps with the expected locus of IPC electrons, with the overlap concentrated on the higher-energy side of the IPC locus, as discussed in detail in Sec.~V.

In all simulated configurations, muons and electrons together account for the bulk of the cosmic-ray response, both in singles and in coincidence. The coincidence requirement specifically suppresses the photon contribution, which falls below $1\%$ in every configuration, while penetrating charged particles are retained. The detailed species fractions vary with orientation and shielding, but this hierarchy is consistent across the configurations studied. The secondary neutron contribution is below $1\%$ in unshielded configurations and rises to several percent inside the lead chamber, attributed to high-energy muon spallation and capture reactions in the lead~\cite{Baginova}.
\subsection{Terrestrial gamma-ray background model}
\label{sec:terrestrial_gamma_model}

The terrestrial gamma-ray background arises primarily from the decay of the long-lived primordial isotopes $^{40}\mathrm{K}$, $^{232}\mathrm{Th}$, and $^{238}\mathrm{U}$ present in the building materials and underlying ground of the laboratory. Modeling this contribution presents a different set of difficulties from the cosmic-ray case. The activities and spatial distributions of these emitters cannot be predicted from first principles, and their absolute intensities depend on the construction history and material composition of each laboratory. A direct measurement of the ambient gamma-ray field is therefore required as input to any quantitative model. The silicon detectors used in this experiment are not suitable for this characterization. Their interaction probability for MeV-scale photons is small, and the resulting spectra are featureless Compton continua that carry no information about the underlying emitter composition. The gamma-ray field was therefore characterized independently using a large-volume NaI(Tl) scintillation detector, whose high-Z active medium and large fiducial volume yield strong photopeak structures from each decay chain. The measured NaI spectrum was then decomposed into contributions from the individual decay chains and used to define a source description that was transferred to the silicon detector response through a dedicated Monte Carlo procedure. The general approach follows the methodology developed in Ref.~\cite{Vepsalainen2020, fowler2024prxq} for gamma background modeling in superconducting quantum circuit substrates, adapted here to present geometry.

\subsubsection{NaI(Tl) measurement and template fit}
\label{sec:nai_template_fit}

\begin{figure}[t]
\centering
\includegraphics[width=1.0\columnwidth]{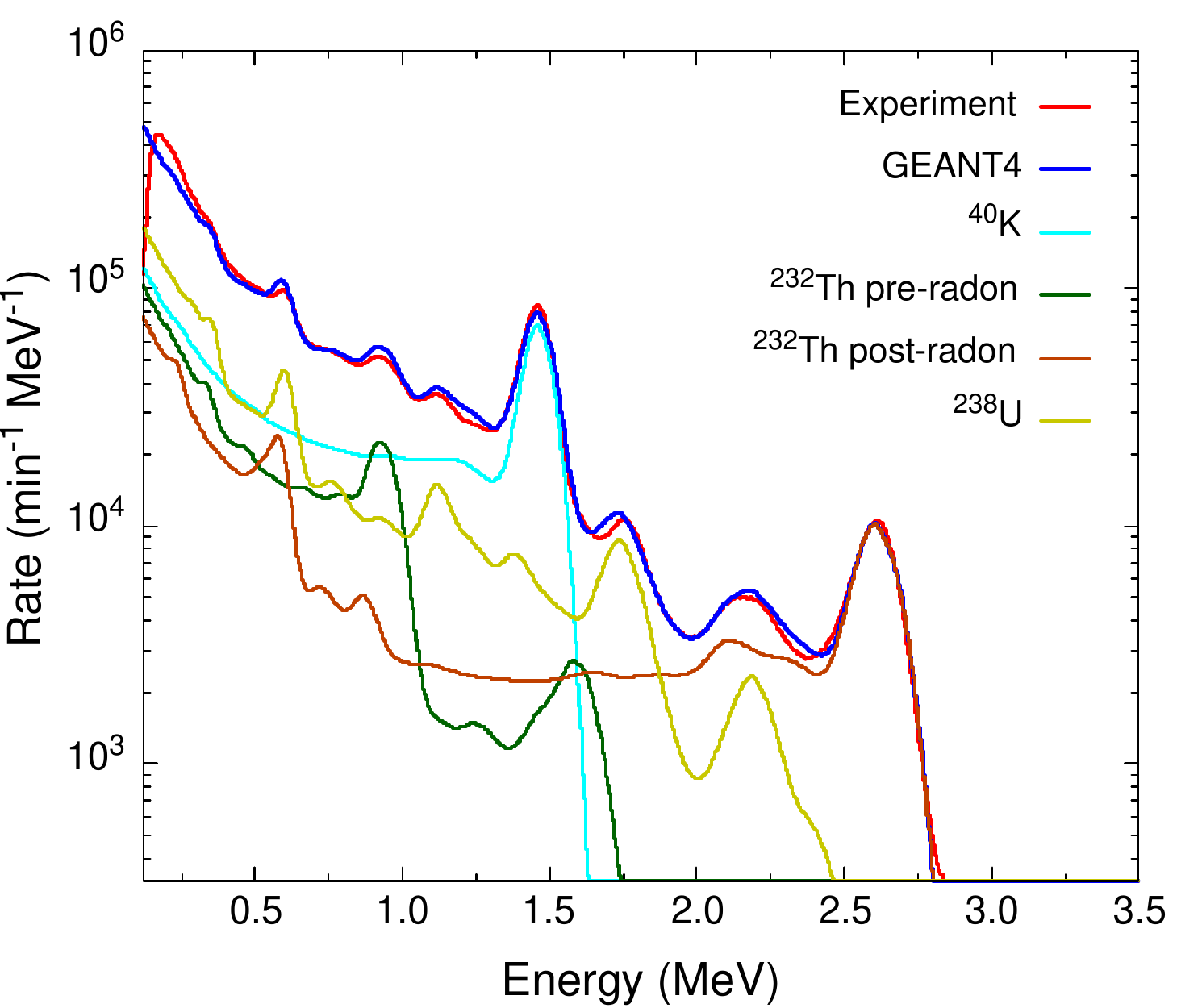}
\caption{Measured NaI(Tl) spectrum in the preparation room together with the GEANT4 template-fit decomposition into contributions from \({}^{40}\mathrm{K}\), the \({}^{238}\mathrm{U}\) chain, and the pre-radon and post-radon segments of the \({}^{232}\mathrm{Th}\) chain.}
\label{fig:nai_spectrum}
\end{figure}

A large-volume $8'' \times 12''$ NaI(Tl) scintillation detector was placed at the same position and height above the floor as the silicon telescope in the preparation room. Background data were accumulated for a large live time of several days to ensure good statistical sensitivity to weak emission lines. The measured spectrum, shown in Fig.~\ref{fig:nai_spectrum}, displays the characteristic gamma lines of the three primordial decay chains, the long-lived isotopes $^{40}\mathrm{K}$, $^{232}\mathrm{Th}$, and $^{238}\mathrm{U}$, which together with their decay daughters dominate the natural gamma-ray background in surface laboratories. The most prominent lines are $^{214}\mathrm{Pb}$ ($352~\mathrm{keV}$), $^{214}\mathrm{Bi}$ ($609.3$, $1120$, $1764$, and $2204~\mathrm{keV}$), $^{40}\mathrm{K}$ ($1460.8~\mathrm{keV}$), and $^{208}\mathrm{Tl}$ ($2614.5~\mathrm{keV}$).

The measured spectrum was decomposed into contributions from the individual source components using a template-fitting procedure. Four source templates were generated by GEANT4 simulation, corresponding to $^{40}\mathrm{K}$, the full $^{238}\mathrm{U}$ decay chain, and the pre-radon and post-radon segments of the $^{232}\mathrm{Th}$ chain. The $^{232}\mathrm{Th}$ chain was split at the radon stage because $^{220}\mathrm{Rn}$ is a noble gas with a half-life of $55~\mathrm{s}$, capable of partially escaping from surface layers of the source material and breaking secular equilibrium between the pre-radon and post-radon segments. The $^{222}\mathrm{Rn}$ intermediate in the $^{238}\mathrm{U}$ chain has a much longer half-life of $3.8~\mathrm{days}$ but the same physical possibility of escape and in principle the $^{238}\mathrm{U}$ chain could be split analogously. In practice, however, the post-radon segment of $^{238}\mathrm{U}$ produces most of the strongest lines in the measured spectrum, while the pre-radon segment is poorly constrained by the available data. The chain was therefore retained as a single template, with the resulting fitted rate effectively reflecting the post-radon segment.

The simulation geometry consisted of a concrete floor slab of $22~\mathrm{cm}$ thickness with the NaI detector placed above it. Gamma emitters were distributed uniformly throughout the floor volume only. The validity of this floor-only assumption was tested in two ways. First, independent simulations performed with sources placed in the walls and ceiling produced NaI response functions with the same spectral shape but different absolute amplitudes, indicating that the spectral shape carries no useful information about the spatial distribution of the emitters. Second, the measured single-detector gamma response of the silicon telescope was unchanged when the telescope was rotated from horizontal to vertical orientation, consistent with a response shape that is independent of the angular distribution of the incident flux for small silicon volumes. The floor-only assumption therefore captures the correct spectral shape, with the geometric integration absorbed into the fitted emission rates.

For each source template, gamma-emission lines and their relative intensities were taken from the NuDat 3~\cite{NuDat3} nuclear data library and configured as discrete energy spectra in the GEANT4 General Particle Source. Emission positions were sampled uniformly within the concrete slab and emission directions were sampled isotropically. Each simulation was normalized by the number of generated primaries to obtain a detector-response template $T_j(E)$ expressed as detected counts per emitted gamma photon for component $j$. The simulated templates were convolved with an empirical NaI detector-response function before fitting. This function combined a Gaussian core, with energy-dependent width determined from independent calibration measurements using $^{22}\mathrm{Na}$, $^{60}\mathrm{Co}$, and $^{137}\mathrm{Cs}$ sources, and an exponential low-energy tail to model the asymmetric peak shape characteristic of large scintillation detectors. A global resolution-scaling factor and an energy offset were included as nuisance parameters to absorb gain and resolution drift over the long background acquisition.

The expected count rate in the NaI detector was modeled as
\begin{equation}
y_{\mathrm{model}}(E)
=
\sum_{j} R_j t_{\mathrm{live}} \widetilde{T}_j(E),
\label{eq:nai_model}
\end{equation}
where $\widetilde{T}_j(E)$ is the resolution-smeared response template for component $j$, $R_j$ is the corresponding effective gamma emission rate, and $t_{\mathrm{live}}$ is the measurement live time. The emission rates $R_j$ were determined by a non-negative least-squares (NNLS) fit to the measured spectrum over the range $0.3$ to $2.75~\mathrm{MeV}$, with the non-negativity constraint enforcing physically meaningful solutions. The cosmic-ray contribution within this energy range, estimated from the CRY+GEANT4 simulation of Sec.~IV A, was small compared to the terrestrial gamma flux and was subtracted from the measured spectrum prior to fitting. The empirical resolution parameters were treated as global nuisance parameters and optimized simultaneously with the emission rates by minimizing a weighted least-squares objective function. The fitted emission rates $R_j$, represent effective source strengths for the assumed floor-only geometry and are not intended as absolute activity measurements. They are used solely to define the spectral shape and relative component strengths required for the silicon detector response simulation described below.

\subsubsection{Transfer to the silicon detectors}
\label{sec:gamma_transfer_to_silicon}

A direct GEANT4 simulation of the silicon-detector response from the full concrete source volume is computationally impractical. Three effects compound to suppress the useful event yield. Most gamma rays emitted within the concrete are absorbed before reaching the laboratory air. Of those that escape, only a small fraction travel in directions that intersect the silicon telescope, which subtends a small solid angle relative to the extended floor source. The interaction probability per gamma ray in a thin silicon detector is also intrinsically low, since Compton scattering in low-Z silicon is a comparatively weak process at the energies of interest. A brute-force simulation therefore yields negligible statistics in the silicon response even for $10^{10}$ primary histories.

\begin{figure*}[t]
\centering
\includegraphics[width=0.80\textwidth]{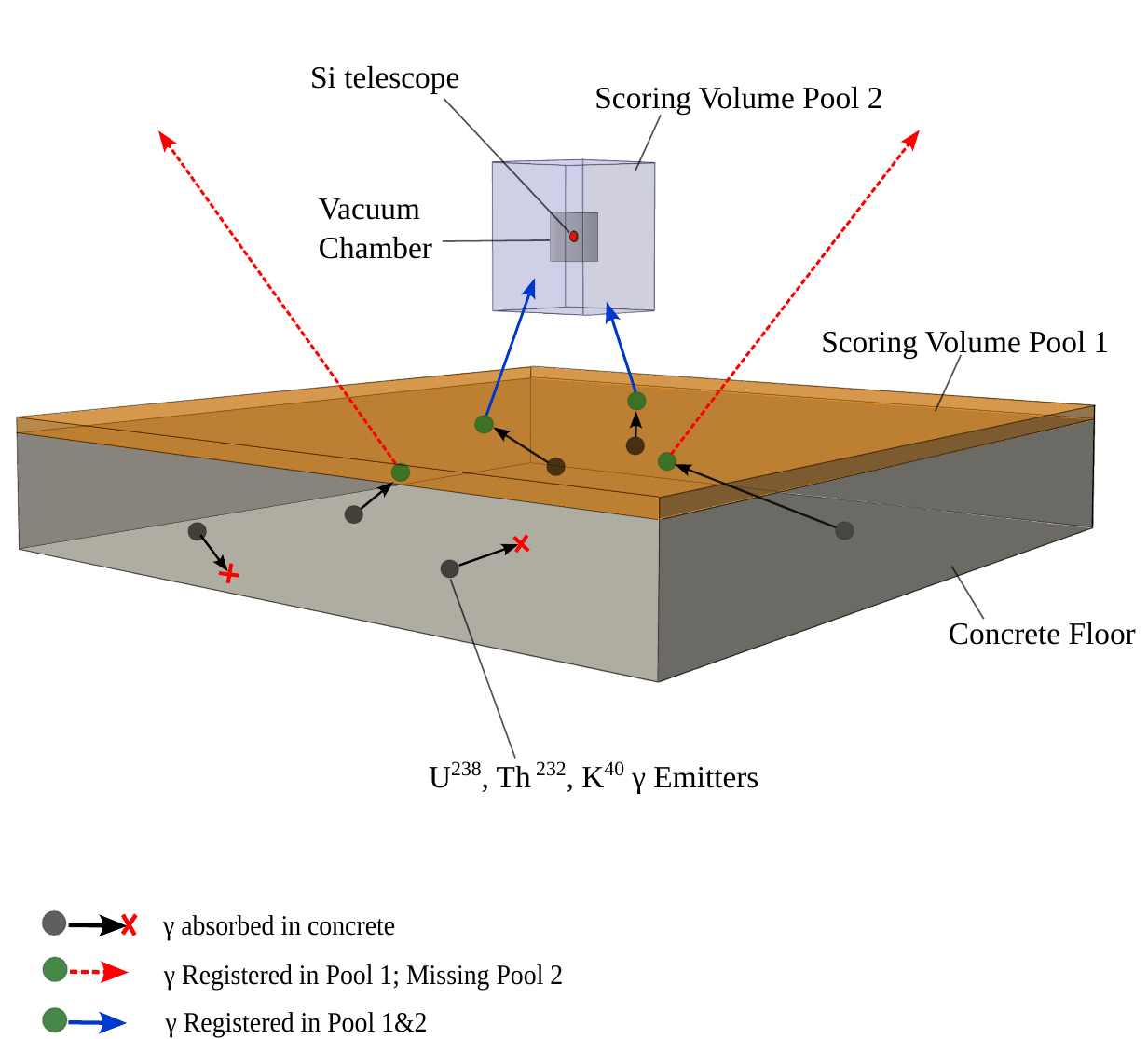}
\caption{Schematic illustration of the two-stage variance-reduction procedure used to transfer the terrestrial gamma source description from the NaI(Tl) template fit to the silicon detector response.}
\label{fig:gamma_variance_reduction}
\end{figure*}

To overcome this limitation, a two-stage variance-reduction procedure based on forward transport with phase-space scoring and conditional importance sampling was implemented\cite{varred1, varred2}. The procedure preserves the full physical transport in the concrete, including all attenuation, scattering, and spectral modification effects, while concentrating the computational effort on gamma rays that are geometrically relevant to the silicon detector. It is illustrated schematically in Fig.~\ref{fig:gamma_variance_reduction}.

In the first stage, gamma rays were generated uniformly throughout a concrete floor slab of $22~\mathrm{cm}$ thickness, using the source compositions and relative emission rates from the NaI template fit. The $22~\mathrm{cm}$ thickness was adopted as a standard approximation. All gamma rays of interest (below $2.615~\mathrm{MeV}$) are attenuated within this depth, since the attenuation length in concrete does not exceed $11~\mathrm{cm}$ in this energy range~\cite{fowler2024prxq,NISTXCOM}. A $0.5~\mathrm{mm}$ thick scoring layer placed immediately above the concrete surface recorded the energy, position, and momentum direction of every gamma ray crossing it, forming the first phase-space pool (Pool 1).

In the second stage, entries in Pool 1 were resampled with a conditional importance-sampling cut, retaining only gamma rays whose extrapolated trajectories intersected a ``hot box'' enclosing both the silicon telescope and the surrounding preparation chamber. The chamber was included in the acceptance volume to capture Compton scattering events near the detector, which contribute to the silicon response through scattered photons and recoil electrons entering the active volume. The cut eliminates gamma rays that would otherwise miss this region entirely while preserving the angular and spectral distribution of those that intersect, which were then transported through the full detector geometry. For each component $j$, the energy deposited in the silicon detectors was recorded for every accepted history and weighted by
\begin{equation}
w_j
=
\frac{R_j \cdot P_{\mathrm{esc},j} \cdot P_{\mathrm{box},j}}
{N_{\mathrm{prim},j}},
\label{eq:gamma_weight}
\end{equation}
where $R_j$ is the effective emission rate from the NaI fit, $P_{\mathrm{esc},j}$ and $P_{\mathrm{box},j}$ are the concrete-escape and box-selection probabilities, and $N_{\mathrm{prim},j}$ is the number of primaries sampled in the second stage. This weight recovers the correct absolute count rate. The total terrestrial gamma response function is
\begin{equation}
G(E)
=
\sum_j Y_j(E),
\label{eq:gamma_response}
\end{equation}
where $Y_j(E)$ is the weighted silicon energy-deposition spectrum for component $j$. Convergence was verified by repeating the second stage with different sample sizes while keeping Pool 1 fixed. After weighting, the resulting spectra agreed in both total rate and spectral shape, confirming that the finite size of Pool 1 introduces no visible artifacts from phase-space recycling.

\subsubsection{Simulated terrestrial gamma response}
\label{sec:simulated_terrestrial_gamma_response}

\begin{figure*}[t]
\centering
\includegraphics[width=0.95\textwidth]{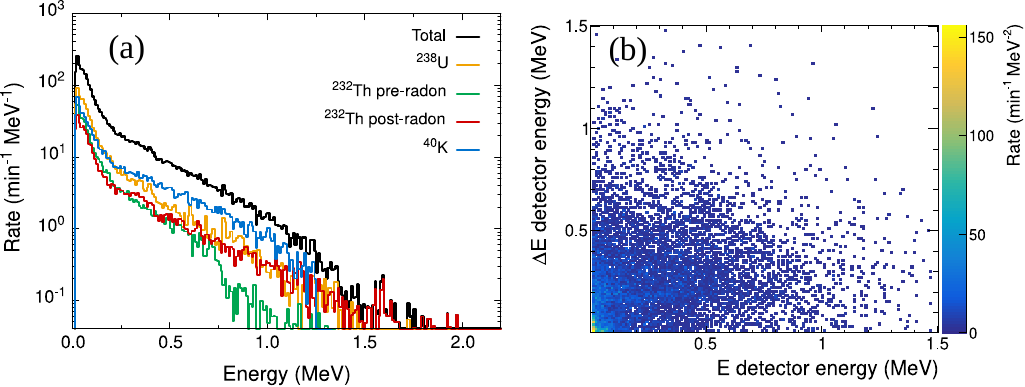}
\caption{Simulated terrestrial gamma-ray response of the silicon telescope in the preparation-room geometry. Panel (a) shows the singles response of the $1~\mathrm{mm}$ $E$ detector, decomposed into contributions from $^{40}\mathrm{K}$, $^{238}\mathrm{U}$, and the pre- and post-radon segments of $^{232}\mathrm{Th}$. Panel (b) shows the corresponding two-dimensional $\Delta E$--$E$ coincidence response.}
\label{fig:si_gamma_response}
\end{figure*}

Figure~\ref{fig:si_gamma_response} shows the simulated terrestrial gamma-ray response in the $1~\mathrm{mm}$ $E$ detector singles spectrum [panel (a)] and in the $\Delta E$--$E$ coincidence histogram [panel (b)]. The singles spectrum is dominated by Compton scattering. It rises sharply below $220~\mathrm{keV}$ and forms an extended continuum that reaches up to about $1.5~\mathrm{MeV}$, falling off rapidly at higher energies. The 2D coincidence response is concentrated at low deposited energies in both detectors, with most of the intensity at $E \lesssim 0.4~\mathrm{MeV}$ and $\Delta E \lesssim 0.4~\mathrm{MeV}$. The distribution is asymmetric with a horizontal band of events extending toward higher $E$-detector energies while $\Delta E$ deposits remain confined to lower values. Events with deposits above $1~\mathrm{MeV}$ are rare in either detector.

\begin{table*}[t]
\caption{
Temporal stability of the measured singles and coincidence background rates.
Rates are given in counts/min. IFV denotes the intrinsic fractional variation,
eq. \ref{eq:f_excess}.}

\label{tab:background_stability}
\centering
\setlength{\tabcolsep}{4pt}
\renewcommand{\arraystretch}{1.15}

\begin{tabular*}{\textwidth}{@{\extracolsep{\fill}}lccccccc}
\toprule
Configuration
& Live time
& \multicolumn{3}{c}{Singles}
& \multicolumn{3}{c}{Coincidence} \\
\cmidrule(lr){3-5}\cmidrule(lr){6-8}
&
& Rate
& $\chi_{\nu}^{2}$
& IFV
& Rate
& $\chi_{\nu}^{2}$
& IFV \\
&
& (min$^{-1}$)
&
& (\%)
& (min$^{-1}$)
&
& (\%) \\
\midrule

Vertical, unshielded
& $286~\mathrm{h}~56~\mathrm{min}$
& $8.57 \pm 0.02$
& $15.2$
& $2.9 \pm 0.3$
& $0.66 \pm 0.01$
& $2.1$
& $3.2 \pm 1.2$ \\

Horizontal, unshielded
& $337~\mathrm{h}~3~\mathrm{min}$
& $6.35 \pm 0.02$
& $10.6$
& $7.2 \pm 0.3$
& $2.01 \pm 0.01$
& $1.8$
& $1.3 \pm 0.8$ \\

Vertical, shielded
& $246~\mathrm{h}~37~\mathrm{min}$
& $2.38 \pm 0.01$
& $10.6$
& $4.7 \pm 0.6$
& $0.299 \pm 0.004$
& $6.1$
& $9.3 \pm 1.7$ \\

Horizontal, shielded
& $327~\mathrm{h}~56~\mathrm{min}$
& $4.00 \pm 0.01$
& $24.4$
& $4.6 \pm 0.4$
& $1.45 \pm 0.01$
& $0.4$
& -- \\

$45^{\circ}$, accelerator chamber
& $397~\mathrm{h}~33~\mathrm{min}$
& $6.89 \pm 0.02$
& $11.8$
& $2.6 \pm 0.4$
& $0.511 \pm 0.004^{(a)}$
& $0.9$
& -- \\

$45^{\circ}$, accelerator chamber, TUKAN$^{(b)}$
& $515~\mathrm{h}~13~\mathrm{min}$
& $4.51 \pm 0.01$
& $10.2$
& $3.2 \pm 0.4$
& --
& --
& -- \\

\bottomrule
\end{tabular*}

\vspace{2mm}
\begin{minipage}{\textwidth}
\footnotesize
The unshielded and shielded preparation-room measurements were performed outside
and inside the lead chamber, respectively.
$^{(a)}$The accelerator-chamber coincidence values were evaluated using the higher-threshold
coincidence gate required for the beam-measurement configuration. They are therefore
not directly comparable to the preparation-room coincidence values.
$^{(b)}$TUKAN measurement with a separate $1~\mathrm{mm}$, $150~\mathrm{mm}^{2}$ PIPS detector and an independent analog readout chain.
\end{minipage}

\end{table*}

\subsection{Background model validation and stability}
\label{sec:background_validation_stability}

The cosmic-ray and terrestrial gamma response functions developed in Sections IV A and IV B were combined to form a total background model. The model was compared directly against the singles spectra measured during long background runs in the preparation room, with the detector telescope in different orientations and shielding configurations, and in the accelerator chamber in the beam-measurement geometry. This subsection presents the validation of the combined model, examines how the background composition depends on detector orientation and shielding, and quantifies the temporal stability of the singles and coincidence response functions.

\subsubsection{Validation of the combined model}
\label{sec:combined_model_validation}

The simulated cosmic-ray and terrestrial-gamma response functions were normalized to the experimental live time, as described in the previous subsections. The integrated rates and component contributions reported in this section are evaluated from $260~\mathrm{keV}$ to $1.3~\mathrm{MeV}$ for the singles spectra in all configurations. For the preparation-room coincidence measurements, the 2D gate shown in  Figure~\ref{fig:coincidence_orientation_2panel} was used. In the accelerator chamber, increased electronic noise in the $\Delta E$ detector required a higher practical threshold and therefore a different 2D gate, as discussed in Sec.~V. These analysis windows were chosen to cover the relevant energy regions where the IPC response is expected from simulations and where the background model must be quantitatively constrained.

The combined model reproduces the measured singles spectral shape in all configurations after allowing for a small global amplitude correction. This correction accounts for residual normalization uncertainties in the simulation framework, including approximations in the room and ceiling geometry, the $22~\mathrm{cm}$ concrete floor-thickness proxy, the uncertainty inherited from the NaI spectral fit, and the finite accuracy of the CRY cosmic-ray flux normalization. To absorb this offset, a single nuisance parameter $k_{\mathrm{amp}}$ was introduced as a global amplitude correction factor, so that the model prediction for the measured singles spectrum becomes
\begin{equation}
M(E)
=
k_{\mathrm{amp}}\left[C(E)+G(E)\right].
\label{eq:background_model_kamp}
\end{equation}
A single shared factor was used rather than two independent factors for $C(E)$ and $G(E)$ because the spectral shapes of both cosmic and terrestrial gamma components are strongly degenerate over the fitting range. A two-parameter fit yields a correlation coefficient $\rho=-0.91$ in the covariance matrix, confirming that the cosmic and terrestrial gamma amplitudes cannot be independently constrained from the spectral data alone. The factor $k_{\mathrm{amp}}$ was determined for each measured background run by minimising the Poisson negative log-likelihood (NLL) objective function. Within the present statistical precision, no additional energy-dependent shape correction was required to reproduce the measured spectral response, confirming that the combined model captures the dominant background components correctly. The simulated 2D $\Delta E$--$E$ coincidence response functions were scaled to the experimental live time in the same way as the singles response functions but without applying an amplitude correction, and are used as a live-time-normalised reference for the spectral shape and component composition of the coincidence background rather than for an absolute-rate comparison.

\begin{figure}[t]
\centering
\includegraphics[width=0.95\columnwidth]{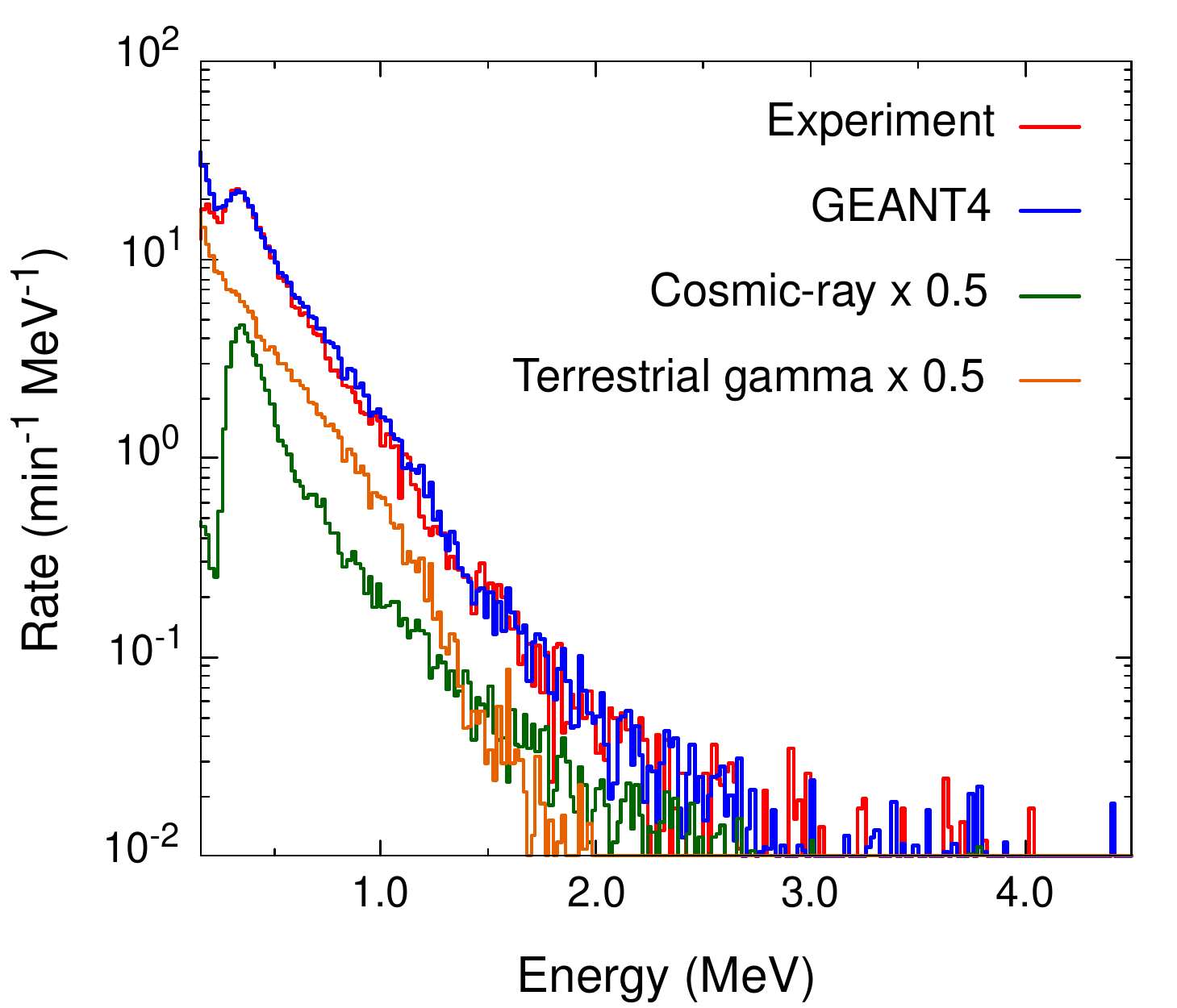}
\caption{
Horizontal-orientation energy spectrum compared with the GEANT4 model. Individual components are scaled by 0.5 for visibility
}
\label{fig:horizontal_validation_spectrum}
\end{figure}

Figure~\ref{fig:horizontal_validation_spectrum} shows the experimental singles spectrum measured in the horizontal unshielded configuration overlaid with the total simulation after application of $k_{\mathrm{amp}}$, decomposed into the cosmic-ray and terrestrial-gamma contributions. The decomposition shows that the terrestrial-gamma component dominates the singles spectrum in this configuration, particularly below $1~\mathrm{MeV}$.

A physically significant result emerges from comparing the terrestrial-gamma to cosmic-ray ratios between the 1D and 2D spectra. In the 1D singles spectrum, the integrated cosmic-ray and terrestrial-gamma contributions were obtained from the combined model fitted to the experimental spectrum by NLL minimization, and the terrestrial-gamma contribution exceeds the cosmic-ray contribution by a factor of $3.23$ above $170~\mathrm{keV}$. In the 2D $\Delta E$--$E$ coincidence histogram, with $C_{\mathrm{2D}}(\Delta E,E)$ and $G_{\mathrm{2D}}(\Delta E,E)$ scaled to the experimental live time, this ratio inverts to $0.67$, with the cosmic-ray component now dominant. The inversion occurs because gamma rays interact predominantly through isolated Compton events in a single detector and are suppressed by the coincidence requirement. The terrestrial gamma component is considered the primary contributor of temporal fluctuations in the singles spectrum, as discussed in Sec.~IV C 3, its suppression in coincidence mode is the main mechanism responsible for the improved stability of the 2D observable.

\subsubsection{Orientation and shielding dependence}
\label{sec:orientation_shielding_dependence}

The amplitude correction factor $k_{\mathrm{amp}}$ has different mean values across different configurations. In the unshielded preparation room, $k_{\mathrm{amp}}=0.871$ in the horizontal orientation and $k_{\mathrm{amp}}=1.231$ in the vertical orientation. The simulation therefore slightly overestimates the horizontal singles rate and underestimates the vertical one. This difference is consistent with the floor-only approximation used in the terrestrial-gamma simulation. In the horizontal orientation, the detector face is directed toward the floor and ceiling, where the floor-only source description captures most of the relevant flux. But in the vertical orientation, the detector face is exposed to a lateral gamma flux from the walls that is absent from the simulated source distribution. The required upward correction in the vertical case is consistent with this missing wall contribution. Inside the accelerator chamber, with the telescope in the beam-measurement geometry, $k_{\mathrm{amp}}\simeq 1.4$, reflecting a different local material distribution and radiation field surrounding the detector.

The mean singles count rates summarised in Table II show how the background composition varies with geometry and shielding. In the preparation room, the unshielded vertical configuration shows the highest singles rate at $8.57 \pm 0.02~\mathrm{cpm}$, while the unshielded horizontal configuration gives $6.35 \pm 0.02~\mathrm{cpm}$. Inside the lead chamber, the singles rates fall to $2.28 \pm 0.01~\mathrm{cpm}$ and $4.00 \pm 0.01~\mathrm{cpm}$ respectively. The reduction produced by lead shielding is markedly different between the two orientations: $72\%$ in vertical orientation but only $37\%$ in horizontal orientation. This difference is consistent with the vertical singles spectrum being dominated by terrestrial gamma radiation, which is efficiently attenuated by $11~\mathrm{cm}$ of lead shielding, while the horizontal spectrum contains a substantially larger fraction of penetrating cosmic-ray events that are not appreciably shielded by lead of this thickness.

\begin{figure*}[t]
\centering
\includegraphics[width=0.95\textwidth]{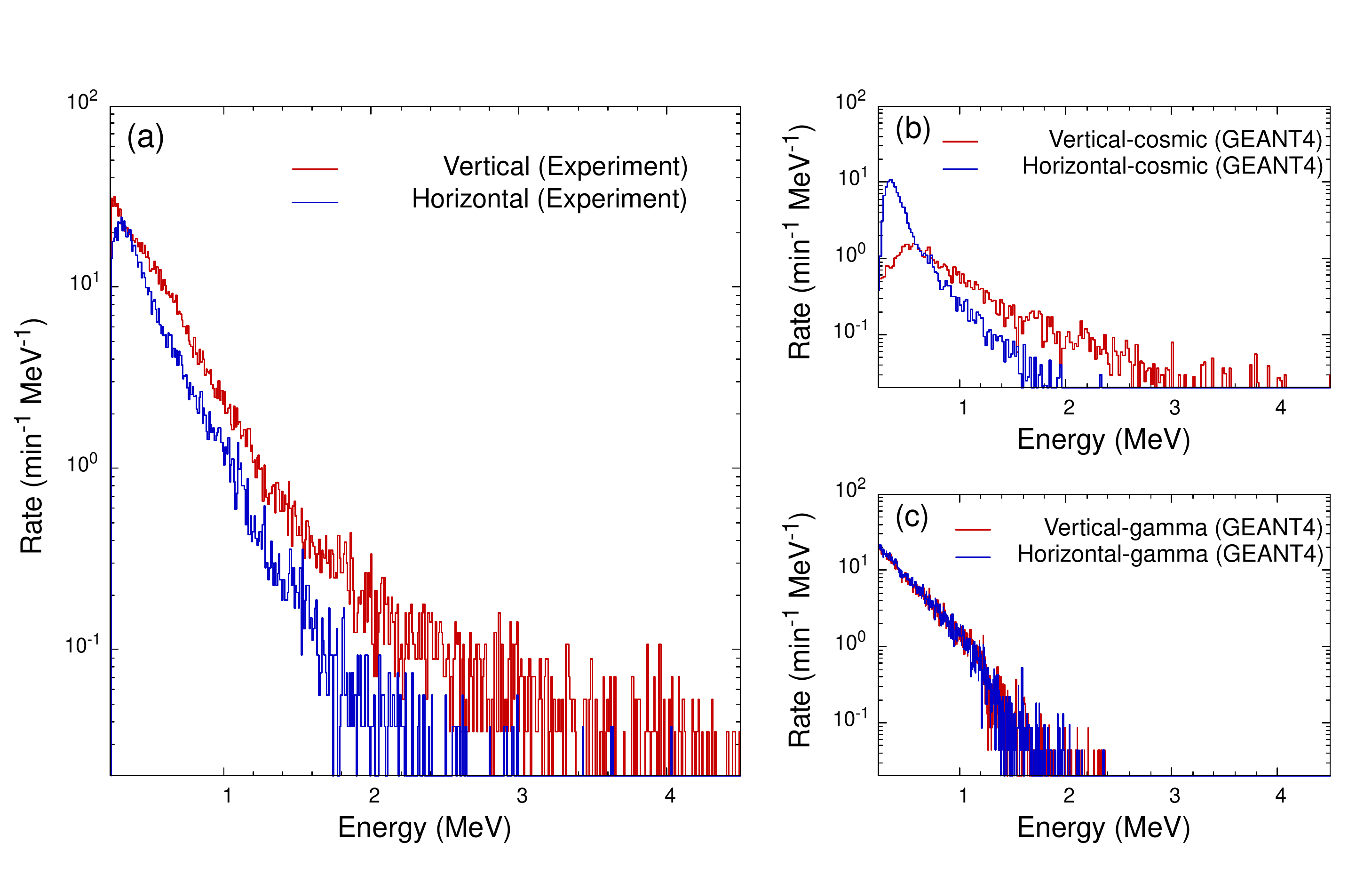}
\caption{
Orientation dependence of the measured and simulated preparation-room spectra. Panel (a) compares the vertical and horizontal experimental singles spectra. Panels (b) and (c) show the corresponding GEANT4 cosmic-ray and terrestrial gamma-ray response functions, respectively. 
}
\label{fig:orientation_spectra}
\end{figure*}

Figure~\ref{fig:orientation_spectra} shows the measured horizontal and vertical singles spectra in the preparation room, alongside the corresponding GEANT4-simulated cosmic-ray and terrestrial gamma response functions for each orientation. The two terrestrial gamma response functions are nearly identical in shape, as expected from the floor-only validation in Sec.~IV B. The cosmic-ray response functions, however, differ substantially between orientations. The vertical orientation selects a population of more oblique muons whose longer path lengths through both detectors produce a characteristic broadening and redistribution of the energy deposition. The shape difference between the measured spectra is therefore attributable primarily to the orientation-dependent cosmic-ray component, while the terrestrial gamma contribution provides an orientation-independent spectral baseline whose amplitude varies with geometry.

\begin{figure*}[t]
\centering
\includegraphics[width=0.90\textwidth]{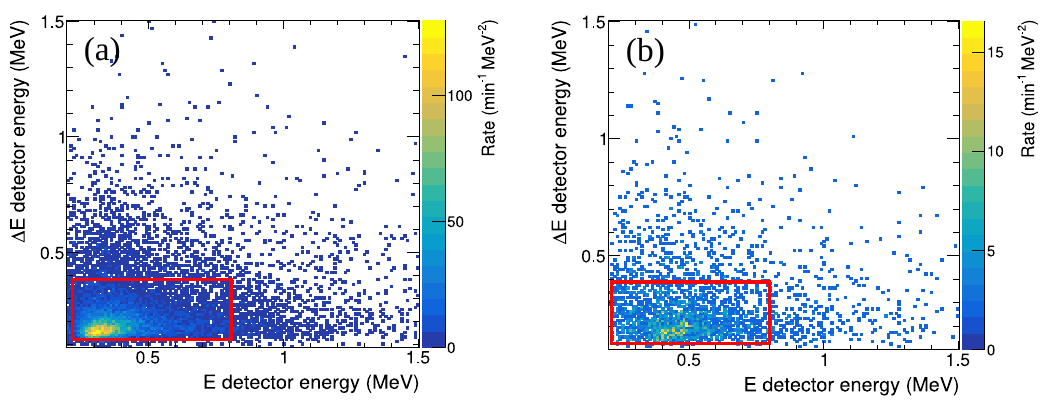}
\caption{
Experimental $\Delta E$--$E$ coincidence spectra measured in the preparation room for the  (a) horizontal and (b) vertical detector orientations. 
}
\label{fig:coincidence_orientation_2panel}
\end{figure*}

The orientation dependence is even more pronounced in the 2D coincidence count rates. Figure~\ref{fig:coincidence_orientation_2panel} shows the experimental $\Delta E$--$E$ coincidence histograms for horizontal and vertical orientations in the preparation room. The horizontal configuration outside lead gives a 2D rate of $2.01 \pm 0.01~\mathrm{cpm}$, while the vertical configuration outside lead gives only $0.66 \pm 0.01~\mathrm{cpm}$, a reduction of $67\%$. Inside the lead chamber, the horizontal 2D rate is $1.45 \pm 0.01~\mathrm{cpm}$ and the vertical is $0.299 \pm 0.004~\mathrm{cpm}$, a reduction of $79\%$. This 2D ordering reflects the geometric efficiency of the detector telescope for downward-going muon tracks. In horizontal orientation the two stacked detectors lie along the dominant muon direction, maximising the coincidence rate, while in vertical orientation the detectors present a smaller cross section to the downward muon flux. This implies that tilting the telescope toward the vertical is an effective, passive measure for reducing the 2D coincidence background, exploiting the angular structure of the cosmic-ray flux at ground level.

Inside the accelerator chamber, with the telescope installed in the beam-measurement geometry, the singles count rate is $6.89 \pm 0.02~\mathrm{cpm}$. This configuration is the one used for the IPC search, and its temporal stability is examined in Sec.~IV C 3.

\subsubsection{Temporal stability of the background response}
\label{sec:background_temporal_stability}

The temporal stability of the background response was characterised using long-duration measurements in each configuration. The total live times are listed in Table II. Each measurement was divided into sub-runs of approximately equal duration, ranging from $23$ to $41~\mathrm{h}$, yielding 7 to 11 sub-runs per configuration. For each configuration, the singles and coincidence count rates were analysed using two complementary stability metrics, the reduced chi-squared statistic $\chi_{\nu}^{2}$ and the intrinsic fractional variability $F_{\mathrm{excess}}$. The former quantifies the statistical significance of the observed run-to-run scatter relative to counting statistics, whereas the latter estimates the fractional amplitude of any residual non-statistical variation \cite{IVF}.

The reduced chi-squared statistic is defined as
\begin{equation}
\chi_{\nu}^{2}
=
\frac{1}{N-1}
\sum_{i=1}^{N}
\frac{\left(R_i-\bar{R}\right)^2}{\sigma_i^2},
\label{eq:reduced_chi_squared}
\end{equation}
where $R_i$ is the count rate of sub-run $i$, $\sigma_i$ is its statistical uncertainty, $\bar{R}$ is the weighted mean rate, and $N$ is the number of sub-runs. The values substantially above unity indicate a scatter inconsistent with statistical fluctuations alone, but do not by themselves quantify the amplitude of the underlying physical variation. To estimate this amplitude, the intrinsic fractional variability was defined as:
\begin{equation}
F_{\mathrm{excess}}
=
\frac{1}{\bar{R}}
\sqrt{\max\left(0,S^2-\langle\sigma^2\rangle\right)}
\times 100\%,
\label{eq:f_excess}
\end{equation}
where $S^2$ is the sample variance of the sub-run count rates and $\langle\sigma^2\rangle$ is the mean squared statistical uncertainty. This quantity estimates the non-statistical component of the run-to-run scatter after subtracting the contribution expected from counting statistics. When $S^2 \leq \langle\sigma^2\rangle$, the observed scatter is compatible with statistical fluctuations and $F_{\mathrm{excess}}$ is reported as zero. The Uncertainties on $F_{\mathrm{excess}}$ were obtained by Monte Carlo propagation, generating synthetic datasets in which each $R_i$ was fluctuated within its Poisson uncertainty and $F_{\mathrm{excess}}$ recomputed. This approach was preferred over analytic propagation because $F_{\mathrm{excess}}$ is a nonlinear estimator bounded below by zero, with an asymmetric distribution when the excess variance is small or when the number of sub-runs is limited.

The stability metrics are summarised in Table II. In the singles spectra, $\chi_{\nu}^{2}$ exceeds unity in every configuration, ranging from $10.2$ to $24.4$ in the preparation room and $11.8$ in the accelerator chamber. The intrinsic fractional variability $F_{\mathrm{excess}}$ ranges from $2.6\%$ to $7.2\%$, with the highest value in the horizontal unshielded preparation room and the lowest in the accelerator chamber. The singles background response is therefore non-stationary in every geometry studied, with real run-to-run variations at the few-percent level.

The coincidence response shows substantially better temporal stability in the horizontal configurations. In the shielded measurement, $F_{\mathrm{excess}}=0$, with no resolved excess variability beyond counting statistics. The outside-lead measurement gives only a small residual non-statistical variation, $F_{\mathrm{excess}}=1.3\pm0.8\%$, much smaller than in the corresponding singles spectrum. The improvement follows naturally from the gamma-to-cosmic ratio inversion established in Sec.~IV C 1. By suppressing the dominant terrestrial-gamma component, the coincidence requirement removes much of the physically variable contribution and leaves a background response dominated by the more stable cosmic-ray component. The vertical 2D configurations are an exception. The shielded vertical configuration gives $F_{\mathrm{excess}}=9.3\pm1.7\%$ while the unshielded vertical configuration gives $F_{\mathrm{excess}}=3.2\pm1.2\%$, both higher than their horizontal counterparts. This enhanced variability is consistent with the angular-selection effect discussed in Sec.~IV B. In the vertical orientation, the narrow geometric cross section of the detectors suppresses near-vertical muons and preferentially accepts more oblique tracks at larger zenith angles. Such tracks are expected to be more sensitive to atmospheric temperature and pressure variations that modulate meson decay in the upper atmosphere \cite{castelon, Muonvariation}. Inside the accelerator chamber, the raised $\Delta E$ threshold required a different coincidence gate, so the accelerator coincidence rate is not directly comparable to the preparation-room values. Within the gate used for the IPC signal extraction in Sec.~\ref{sec:signal}, $F_{\mathrm{excess}}=0$ and $\chi_{\nu}^{2}=0.9$, indicating that no excess run-to-run variability is resolved beyond counting statistics.

As a cross-check of possible instrumental contributions to the observed singles fluctuations, an independent measurement was performed inside the accelerator chamber using a $1~\mathrm{mm}$-thick PIPS detector with a $150~\mathrm{mm}^{2}$ sensitive area. The detector was read out through a separate analogue electronics chain and a TUKAN 8K multichannel analyser. This system gives $F_{\mathrm{excess}}=3.2\pm0.4\%$ in singles, in agreement with the $F_{\mathrm{excess}}=2.6\pm0.4\%$ obtained with the $1~\mathrm{mm}$-thick, $300~\mathrm{mm}^{2}$ PIPS detector read out by the CAEN digitiser. Since the two measurements used different detectors, preamplifiers, amplifiers, and multichannel analysers, the agreement indicates that the observed singles fluctuations are not an artefact of the front-end electronics or digitiser, but instead reflect real variations in the local radiation environment.

The $k_{\mathrm{amp}}$ values obtained from the long beam-off measurements in the accelerator chamber define the systematic uncertainty envelope on the background normalisation used in the signal extraction of Sec.~\ref{sec:signal}, and the stability of the coincidence observable supports its use as a quantitative observable subject to the threshold limitations described above.
\section{Application to the IPC Channel Search}
\label{sec:signal}

The validated background model and the amplitude correction envelope developed in Sec.~IV are now applied to the search for the IPC channel in low-energy DD fusion. The 1 mm \(E\)-detector singles spectrum recorded at a deuteron beam energy of 18 keV provides the primary observable from which the electron-to-proton branching ratio is extracted. The 2D \(\Delta E\)--\(E\) coincidence response, recorded over the full 10--20 keV beam-energy range, is treated separately in Sec.~V D as a threshold-limited cross-check.

\subsection{Beam-on signal-plus-background model}
\label{sec:beam_on_model}

Under beam-on conditions, the silicon telescope records charged DD reaction products in addition to the natural-radiation background characterized in Sec.~IV. The protons, tritons, and \(^{3}\)He ions emitted from the target are stopped in the 500 \(\mu\)m \(\Delta E\) detector and do not reach the 1 mm \(E\) detector. The 1 mm \(E\)-detector spectrum therefore receives contributions from four physical components: the cosmic-ray response \(C(E)\) and the terrestrial gamma response \(G(E)\) characterized in Sec.~IV; the response \(N(E)\) to the 2.45 MeV neutrons from the \(^{2}\mathrm{H}(d,n)^{3}\mathrm{He}\) branch; and the response \(S_{\mathrm{IPC}}(E)\) to the high-energy electron-positron pairs from the proposed IPC decay of the near-threshold \(0^{+}\) resonance in \(^{4}\)He. The expected spectrum in the 1 mm \(E\) detector is therefore modelled as
\begin{equation}
M(E) = k_{\mathrm{amp}}\,[C(E) + G(E)] + N(E) + S_{\mathrm{IPC}}(E),
\label{eq:beam_on_model}
\end{equation}
where \(C(E)\), \(G(E)\), and \(N(E)\) are normalized to the experimental live time, and \(S_{\mathrm{IPC}}(E)\) carries the branching ratio \(\mathrm{BR}\) as the only free parameter of the analysis. The fit is performed over the interval \(0.26 \leq E \leq 1.3\) MeV, which contains the broad partial-absorption response of the IPC \(e^{+}e^{-}\) in the 1 mm detector as shown in Figure \ref{fig:spectrum_18kev}. The validated laboratory background model over this interval from Sec.~IV can be directly used in the beam-on fit without further extrapolation.

A single shared amplitude correction factor \(k_{\mathrm{amp}}\) multiplies the combined laboratory background term \(C(E)+G(E)\). As established in Sec.~IV C, the cosmic-ray and terrestrial-gamma spectral shapes are strongly degenerate over this fit window, and a single shared factor is therefore the only reliable parameterization. The factor \(k_{\mathrm{amp}}\) is not floated as a free parameter in the beam-on fit. Its admissible range is instead constrained externally from the long-duration beam-off background measurements in the accelerator chamber reported in Sec.~IV C 3. From these measurements, \(k_{\mathrm{amp}}\) is summarized by three discrete values
\begin{equation}
k_{\mathrm{amp}} \in \{k_{\min},\, \bar{k},\, k_{\max}\},
\label{eq:k_envelope}
\end{equation}
corresponding to the lowest, time-weighted mean, and highest values observed over the beam-off campaign. Treating these three values as fixed background normalizations propagates the run-to-run background variability directly into the branching-ratio analysis as a systematic envelope, rather than collapsing it onto a single free parameter with no empirical bound.

\subsection{Neutron and IPC response function}
\label{sec:neutron_ipc_response}

\subsubsection{Neutron response}
\label{sec:neutron_response}

The 2.45 MeV neutrons emitted in the \(^{2}\mathrm{H}(d,n)^{3}\mathrm{He}\) branch of the DD reaction lie in the fast-neutron regime, where the silicon telescope response is dominated by elastic and inelastic scattering. Radiative capture cross sections on the relevant structural nuclei such as \(^{56}\)Fe are smaller by approximately two orders of magnitude at this energy \cite{neutroncapture} and can be neglected. The elastic contribution arises from silicon nuclear recoils produced by neutron scattering inside the active detector volume, since recoil ranges are too short to permit appreciable contributions from material outside the detector. The inelastic contribution arises predominantly from prompt gamma rays emitted by the residual nucleus following inelastic scattering on surrounding materials \cite{promptgamma1, promptgamma2}, which then enter the silicon detector and deposit energy through Compton scattering.

The neutron response function \(N(E)\) was generated by a dedicated GEANT4 simulation of the beam-on geometry, comprising the palladium target, molybdenum target holder, aluminum absorber foil, ultra-high-vacuum chamber walls, and the full \(\Delta E\)--\(E\) telescope. A total of \(10^{8}\) monoenergetic 2.45 MeV neutrons were emitted isotropically from the target spot, and the resulting energy depositions in the active silicon volumes were recorded. To attribute each silicon energy deposit to its underlying neutron interaction, an ancestry-based tagging procedure was implemented in the GEANT4 stepping action. The neutron interaction process, elastic scattering, inelastic scattering, neutron capture, or other, was identified at the point where the primary neutron interacted, and the tag was propagated to all descendant particles produced in the resulting cascade.

The simulated response is consistent with the expected interaction mechanism. Above a 10 keV deposition threshold, elastic interactions account for approximately 58\% of the total energy deposition, inelastic processes for approximately 42\%, and neutron capture for approximately 1\%. The capture contribution can therefore be neglected. The total neutron response is concentrated at low deposited energies and falls rapidly toward higher energies, so that its contribution in the upper part of the fit window is small compared to the laboratory background. The normalized neutron response is shown together with the other components of the beam-on model in Fig.~\ref{fig:spectrum_18kev}.

The absolute normalization of \(N(E)\) is fixed using the experimentally measured proton yield. The total number of protons emitted from the target, \(N_p^{\mathrm{tot}}\), is obtained from the measured proton yield \(N_p^{\mathrm{meas}}\) in the \(\Delta E\) singles spectrum through
\begin{equation}
N_p^{\mathrm{tot}} = \frac{N_p^{\mathrm{meas}}}{\varepsilon_p^{\mathrm{geom}}},
\end{equation}
where \(\varepsilon_p^{\mathrm{geom}}\) is the geometric efficiency of the \(\Delta E\) detector for protons emitted from the target, obtained from a separate GEANT4 simulation. At a deuteron beam energy of 18 keV, the proton and neutron channels of the DD reaction are taken to have equal yield \cite{czerski2024prc, prx2025},  and the total number of neutrons emitted during the run is therefore \(N_n^{\mathrm{tot}}=N_p^{\mathrm{tot}}\). The neutron response function used in the fit is then
\begin{equation}
N(E) = \frac{N_n^{\mathrm{tot}}}{N_n^{\mathrm{sim}}}\, n_{\mathrm{sim}}(E),
\label{eq:neutron_norm}
\end{equation}
where \(n_{\mathrm{sim}}(E)\) is the simulated energy-deposition spectrum and \(N_n^{\mathrm{sim}}\) is the number of neutrons in the Monte Carlo. The neutron component is therefore fully constrained before the fit and is not a free parameter.

\subsubsection{IPC response}
\label{sec:ipc_response}

The IPC response function in the 1 mm \(E\) detector was obtained from the same GEANT4 simulation used to determine the IPC detection efficiency in Sec.~III D. Separate response functions \(S_{e^-}(E)\) and \(S_{e^+}(E)\) were generated for the electron and positron components of the IPC pair. The two spectra are nearly identical in the present detector geometry, and the IPC spectral shape \(T_{\mathrm{IPC}}(E)\) used in the spectral fit is therefore taken as the normalized sum,
\begin{equation}
T_{\mathrm{IPC}}(E) =
\frac{S_{e^-}(E) + S_{e^+}(E)}
{\int [S_{e^-}(E) + S_{e^+}(E)]\,dE},
\label{eq:ipc_template}
\end{equation}
with \(\int T_{\mathrm{IPC}}(E)dE=1\) by construction.

The absolute normalization of the IPC response is done with the help of the same measured proton yield \(N_p^{\mathrm{meas}}\) used to normalize \(N(E)\). The protons are detected in the \(\Delta E\) detector, whereas the IPC response is studied in the 1 mm \(E\) detector. The two detectors subtend slightly different solid angles at the target, and this geometric difference is accounted for by the conditional geometric efficiency \(f(\Delta E \mid E)\) defined in Sec.~III. The proton-normalized reference yield entering the IPC term is therefore
\begin{equation}
N_p^{\mathrm{ref}} =
\frac{N_p^{\mathrm{meas}}}{f(\Delta E \mid E)},
\label{eq:proton_ref}
\end{equation}
and the IPC contribution to the 1 mm \(E\)-detector singles spectrum is
\begin{equation}
S_{\mathrm{IPC}}(E) =
\mathrm{BR} \cdot N_p^{\mathrm{ref}} \cdot T_{\mathrm{IPC}}(E),
\label{eq:ipc_signal}
\end{equation}
where \(\mathrm{BR}\) is the electron-to-proton branching ratio and is the only free parameter of the beam-on fit.

\subsection{Branching-ratio extraction from the 1 mm singles spectrum}
\label{sec:br_extraction}

The 1 mm \(E\)-detector singles spectrum measured at a deuteron beam energy of 18 keV was fitted with the model of Eq.~(\ref{eq:beam_on_model}) over the interval \(0.26 \leq E \leq 1.3\) MeV defined in Sec.~V A. The branching ratio \(\mathrm{BR}\) is the only free parameter of the fit.

The fit was performed by minimizing the Poisson negative log-likelihood \cite{likelihoodpaper}
\begin{equation}
-\ln \mathcal{L}(\mathrm{BR}) =
\sum_i
\left[
M_i(\mathrm{BR}) - n_i
+ n_i \ln \frac{n_i}{M_i(\mathrm{BR})}
\right],
\label{eq:nll}
\end{equation}
where \(n_i\) is the number of measured counts in bin \(i\) and \(M_i(\mathrm{BR})\) is the corresponding model prediction. The \(1\sigma\) statistical interval on \(\mathrm{BR}\) was determined from the profile-likelihood condition \(\Delta(-\ln \mathcal{L})=0.5\).

The fit was carried out separately for each of the three fixed background normalizations defined in Eq.~(\ref{eq:k_envelope}), with values \(k_{\min}=1.34\), \(\bar{k}=1.384\), and \(k_{\max}=1.44\). The best-fit branching ratios are
\begin{align}
\mathrm{BR}(k_{\min}) &= 0.037 \pm 0.002, \\
\mathrm{BR}(\bar{k}) &= 0.033 \pm 0.002, \\
\mathrm{BR}(k_{\max}) &= 0.028 \pm 0.002.
\end{align}

The three values vary by approximately 14\% across the admissible range of \(k_{\mathrm{amp}}\). 

\begin{figure*}[t]
\centering
\includegraphics[width=0.99\textwidth]{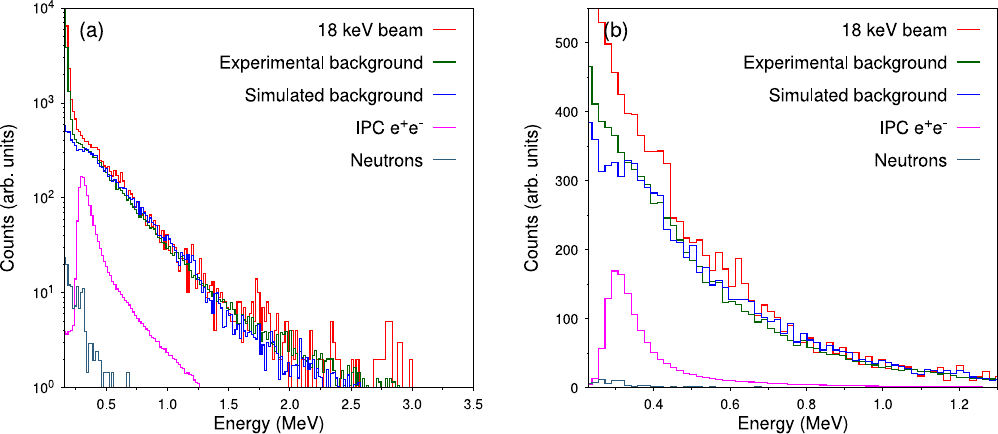}  
\caption{Beam-on singles spectrum measured in the \(1~\mathrm{mm}\) \(E\) detector at \(E_d = 18~\mathrm{keV}\), shown together with the best-fit model at \(\bar{k}\). Panel (a) presents the spectrum on a logarithmic scale over the extended energy range, while panel (b) presents the same spectrum on a linear scale over the fitting interval to provide greater visibility of the enhancement produced by the IPC contribution. The total prediction is decomposed into the experimental and simulated laboratory background, the neutron-induced response, and the IPC electron--positron contribution.}
\label{fig:spectrum_18kev}
\end{figure*}

Taking the central fit at \(\bar{k}\) as the result and the spread between the three fixed-\(k\) solutions as a systematic uncertainty associated with the run-to-run background variability, the branching ratio is
\begin{equation}
\mathrm{BR} =
0.033 \pm 0.002\,(\mathrm{stat.})\,^{+0.003}_{-0.005}\,(\mathrm{syst. envelope}).
\label{eq:BR_result}
\end{equation}
Figure~\ref{fig:spectrum_18kev} shows the measured 18 keV beam-on spectrum together with the best-fit model at \(\bar{k}\), decomposed into the laboratory background, the neutron response, and the IPC contribution.

The null hypothesis \(\mathrm{BR}=0\) is disfavoured for all three background normalizations. The likelihood-ratio significance was computed as
\begin{equation}
Z =
\sqrt{2\,\Delta(-\ln \mathcal{L})_{0}},
\end{equation}
where \(\Delta(-\ln \mathcal{L})_{0}\) is the difference between the negative log-likelihood at \(\mathrm{BR}=0\) and at the best-fit value, evaluated for each fixed \(k_{\mathrm{amp}}\) \cite{likelihoodpaper}. The corresponding significances are \(Z(k_{\min}) = 20.8\sigma\), \(Z(\bar{k}) = 18.5\sigma\), and \(Z(k_{\max}) = 15.7\sigma\). Even for the most conservative background choice, the data require a nonzero IPC-like component with high statistical significance.

\subsection{Two-dimensional coincidence consistency check}

The 2D $\Delta E$--$E$ coincidence histogram provides an independent check of the singles-spectrum result of Sec.~V C. Inside the accelerator chamber, the electronic-noise level was higher than in the preparation room, and stable operation of the telescope required raising the lower-level discriminator thresholds to approximately 295 keV in $\Delta E$ and 220 keV in $E$. The noise level was the same in beam-on and beam-off conditions, indicating that the source is the chamber environment itself, most likely the vacuum pumps, rather than the beam. The simulated IPC partial-absorption peaks are located near 150 keV in the 500 $\mu$m detector and near 315 keV in the 1 mm detector. The $E$-detector threshold therefore preserves the IPC peak in the thick detector, which is the basis for the singles analysis of Sec.~V C. The $\Delta E$ threshold, however, lies at higher energy than the IPC peak in the thin detector, so most of the IPC coincidence events fall below the nominal acceptance of the 2D histogram.

Two factors make a meaningful coincidence test still possible. First, the discriminator turn-on is not perfectly rigid, and a residual population of events is recorded below the nominal threshold. The recorded population decreases progressively at energies further below the threshold, following the standard s-curve shape of a discriminator turn-on whose width is set by the electronic noise on the baseline \cite{Threshold1, Threshold2}. Second, electronic noise in the $\Delta E$ detector is uncorrelated in time with the $E$ detector and is therefore strongly suppressed by the coincidence condition. The coincidence histogram is built with the offline cuts $\Delta E\geq100$ keV, $E\geq220$ keV, and a prompt coincidence window of 210 ns determined from the time-of-flight spectrum using the procedure described in Sec.~III. Under these conditions, the expected number of accidental coincidences from the measured singles rates is approximately 8 counts over the 59246 s of the 18 keV beam run, negligible compared with the gated yields discussed below.

To compensate for the reduced IPC acceptance and to maximise the statistics of the test, the coincidence dataset is summed over the full 10--20 keV beam-energy range in 2 keV steps, for a total live time of 194007 s. The branching ratio extracted from this dataset is therefore an effective average over this energy range rather than a value at a single beam energy.

\begin{figure*}[t!]
\centering
\includegraphics[width=0.99\textwidth]{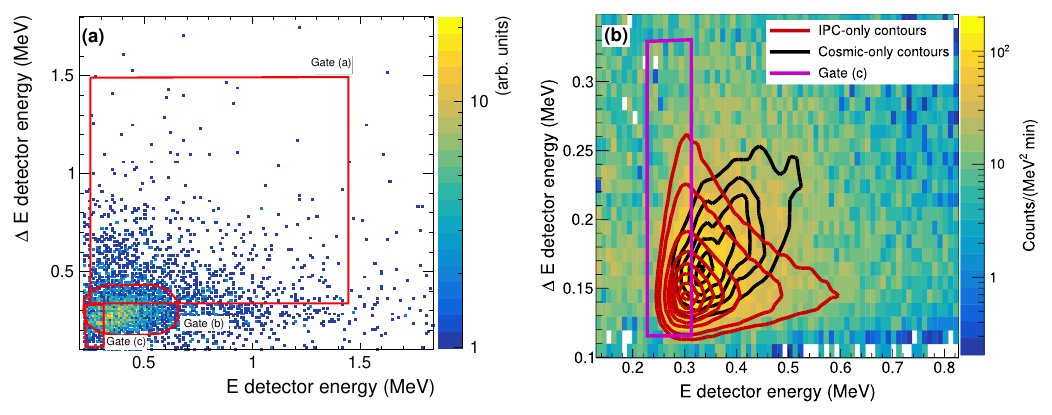}
\caption{Two-dimensional $\Delta E$--$E$ coincidence analysis. (a) Experimental
coincidence histogram summed over the 10--20 keV beam-energy range, with the
three gates overlaid. (b) Simulated IPC (red) and cosmic-ray (black) responses, shown with the contours to compare their spectral shapes and peak positions. The contours span selected levels from 5--95\% and 25--95\% of their respective maxima, and Gate (c) is shown in magenta.}
\label{fig:coincidence_gates}
\end{figure*}

Three gates were defined on the resulting 2D $\Delta E$--$E$ histogram (Fig.~\ref{fig:coincidence_gates}(a)). Gate (a) lies fully above the physical thresholds of both detectors and provides the most direct cross-check, free of any threshold correction. Gate (b) intersects the $\Delta E$ threshold boundary and contains the largest fraction of the observable counts.  Gate (c) lies fully below the $\Delta E$ threshold. Its placement follows from the simulated IPC and cosmic-ray responses overlaid in Fig.~\ref{fig:coincidence_gates}(b), where the IPC response peaks at lower deposited energy than the cosmic-ray response and differs from it in shape. Gate (c) selects the IPC-dominated region while remaining below the peak of the muon-induced cosmic-ray background identified in Sec.~\ref{sec:background}.

The IPC gate efficiency, defined as the fraction of all detectable IPC coincidence events falling inside a given gate, was determined from the simulated IPC response as the ratio of simulated counts inside the gate to the total simulated counts. The resulting efficiencies are $\varepsilon_{2D}^{(a)}=6\%$, $\varepsilon_{2D}^{(b)}=19\%$, and $\varepsilon_{2D}^{(c)}=26\%$. The branching ratio for a given gate is then
\begin{equation}
\mathrm{BR}^{(\mathrm{gate})}
=
\frac{N^{(\mathrm{gate})}_{\mathrm{excess}}}
{\varepsilon^{(\mathrm{gate})}_{2D}\, \varepsilon_{\mathrm{trf}}\, N_{p,\mathrm{tot}}},
\label{eq:BR_gate}
\end{equation}
where $N_{\mathrm{excess}}^{(\mathrm{gate})}$ is the beam-on excess over the experimentally measured chamber background in the same gate, $N_{p,\mathrm{tot}}$ is the total proton yield summed over all beam-energy points, and $\varepsilon_{\mathrm{trf}}$ is a threshold-survival factor that accounts for events lost to the discriminator turn-on.  For gate (a), which lies fully above the physical thresholds, $\varepsilon_{\mathrm{trf}}=1$. For gates (b) and (c), $\varepsilon_{\mathrm{trf}}$ was extracted from the beam-off chamber background. Because gate (a) lies above both thresholds, it is unaffected by the discriminator turn-on and serves as a reference. For a gate $g$,
 \begin{equation} 
\varepsilon_{\mathrm{trf}}^{(g)} = \frac{ \left(N^{(g)}/N^{(a)}\right)_{\mathrm{exp}} }{ \left(N^{(g)}/N^{(a)}\right)_{\mathrm{sim}} }, 
\label{eq:trf} 
\end{equation} 
where $\left(N^{(g)}/N^{(a)}\right)_{\mathrm{exp}}$ and $\left(N^{(g)}/N^{(a)}\right)_{\mathrm{sim}}$ are the counts in gate $g$ relative to gate (a) in the experimentally measured and in the threshold-free GEANT4 background, respectively. The experimentally measured ratio is reduced by the discriminator turn-on, while the simulated ratio is not, so their quotient gives the fraction of gate-$g$ events surviving the discriminator threshold. This yields $\varepsilon_{\mathrm{trf}}^{(b)} = 0.67 \pm 0.05$ and $\varepsilon_{\mathrm{trf}}^{(c)} = 0.59 \pm 0.08$.

Applying Eq.~(\ref{eq:BR_gate}) to each gate gives
\begin{align}
\mathrm{BR}^{(a)} &= 0.011 \pm 0.007, \\
\mathrm{BR}^{(b)} &= 0.008 \pm 0.004, \\
\mathrm{BR}^{(c)} &= 0.004 \pm 0.001.
\end{align}
The uncertainties are dominated by the limited coincidence statistics within each gate and, for gates (b) and (c), by the uncertainty on the threshold-survival factor. The corresponding reference value from the singles analysis is the proton-yield-weighted average of the singles branching ratio over the same 10--20 keV beam-energy range, obtained using the methodology of Sec.~V C at each beam-energy point:
\begin{equation}
\left\langle \mathrm{BR} \right\rangle_{\mathrm{singles}}
=
0.0352
\pm 0.001\,(\mathrm{stat.})
{}^{+0.005}_{-0.007}\,(\mathrm{syst.}).
\label{eq:BR_singles_weighted}
\end{equation}
Each of the three gate values is consistent with the singles-based prediction within combined statistical and systematic uncertainty, with the largest deviation, in gate (c), at approximately the 2$\sigma$ level. The observed variations are attributed mostly to the uncertainty on the model dependent threshold-survival factor used for gates (b) and (c). The 2D coincidence values are therefore consistent with the singles-based expectation, although their statistical reach is too limited to provide an independent quantitative determination of the branching ratio. Therefore the 1 mm singles fit of Sec.~V C remains the primary source of the quantitative result, and the 2D coincidence analysis confirms its internal consistency under the threshold-limited conditions of the present experiment.

\section{Discussion and Conclusion}
\label{sec:discussion}
Previous low-energy DD IPC measurements relied on direct subtraction of an experimentally measured background from the beam-on spectrum. This work replaces that subtraction step with a quantitative, component-by-component background model. Each background component is modelled independently using the GEANT4 Monte Carlo code and validated against long-duration measurements. The observed run-to-run variability in the total background is represented through a single shared amplitude correction factor $k_{\mathrm{amp}}$ whose envelope $\{k_{\min},\bar{k},k_{\max}\}$ is determined from independent long-duration measurements in each detector configuration. The branching ratio is then bracketed by the three solutions corresponding to this envelope, yielding a systematic uncertainty that scales directly with the empirically measured background stability.

\begin{figure}[t]
\centering
\includegraphics[width=\columnwidth]{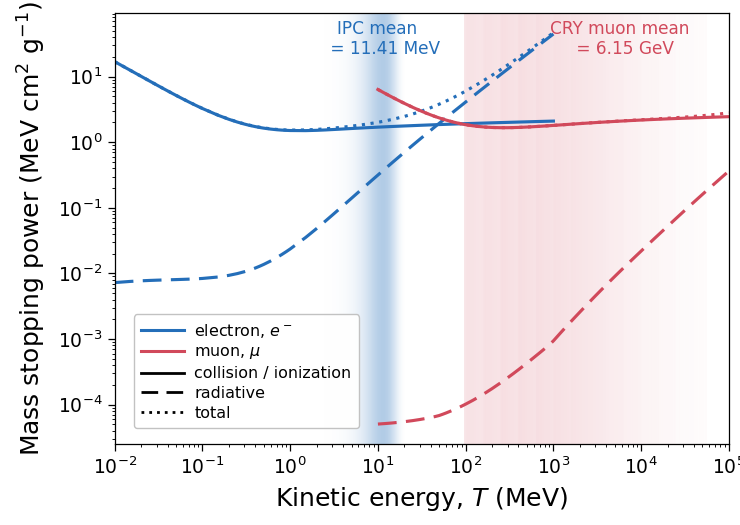}
\caption{Mass stopping powers of electrons and muons in silicon versus kinetic
energy, separated into collisional, radiative, and total components. The blue
and red shading shows the normalized kinetic-energy distributions of the IPC
leptons and cosmic-ray muons, independently normalized, with darker regions
indicating greater relative intensity. Electron and muon values are taken from
NIST ESTAR~\cite{NISTESTAR} and the Particle Data Group tables~\cite{PDG2024},
respectively.}
\label{fig:stopping_power}
\end{figure}

A second result that emerges from comparing the singles and coincidence background characterisations is that the 2D coincidence response is the more stable observable. The singles spectrum is dominated by the terrestrial-gamma background, which exhibits measurable non-statistical day-to-day variations on the few-percent level. The presence of a break in the secular equilibrium of the $^{232}$Th chain at the radon stage, supported by the independent NNLS fits to the NaI(Tl) spectrum in Sec.~IV B, points to radon-progeny variations as a plausible physical mechanism behind the observed instability. The 2D coincidence response suppresses the terrestrial-gamma component, since gamma rays interact predominantly through single Compton scattering in one detector, and the cosmic-ray-correlated component then becomes the dominant residual background. This component is intrinsically more stable than the gamma component, and the resulting 2D background response shows no measurable non-statistical variation in the accelerator-chamber and horizontal configurations. In the beam irradiation geometry, with the telescope at $45^\circ $ to the vertical, the cosmic-correlated background in the $1 \mathrm{mm}$ detector produces a maximum structure between approximately $350$ and $370 \mathrm{keV}$, while the IPC partial-absorption peak lies slightly below this, between $300$ and $315 \mathrm{keV}$. The orientation dependence of the 2D spectral shape is pronounced. In the horizontal orientation, the spectrum carries a strong cosmic-muon-induced structure aligned with the downward muon flux. In the vertical orientation, the detector face presents the smallest geometrical cross section to the dominant downward muon flux, the spectrum is correspondingly diffuse, and the absolute 2D coincidence rate is reduced by 67\% relative to the horizontal orientation. The vertical configuration is therefore the optimal orientation for surface-laboratory weak-particle searches in the same energy regime, with the addition of lead shielding providing a further reduction of the terrestrial-gamma contribution. The present IPC search was performed in the accelerator-chamber geometry without lead shielding. A vertical, lead-shielded coincidence configuration would represent the lowest-background and most stable operating point achievable with the present detector telescope. \\
 The close similarity between the IPC and cosmic-muon responses can be understood from their stopping powers in silicon. Figure~\ref{fig:stopping_power} shows the collisional, radiative, and total mass stopping powers for electrons and muons. At the mean IPC lepton energy of \(11.41~\mathrm{MeV}\), the total electron stopping power is \(2.08~\mathrm{MeV\,cm^{2}\,g^{-1}}\), with the collisional term accounting for approximately \(80\%\) and the radiative term for the remainder. The cosmic-muon spectrum generated by CRY has a mean energy of \(6.15~\mathrm{GeV}\) [Fig.~\ref{fig:cry_input_spectra}]. At this energy, the total muon stopping power is \(2.12~\mathrm{MeV\,cm^{2}\,g^{-1}}\) and is almost entirely collisional. Thus, although the particle energies differ by more than two orders of magnitude, their total stopping powers differ by only about \(2\%\).

In both cases, collisional energy loss is the dominant contribution. Electrons and muons have the same charge magnitude and are highly relativistic, with \(\beta>0.999\), so their collisional stopping powers are of the same order and vary only slowly with energy. The muon nevertheless has the larger collisional contribution, whereas the electron has a much larger radiative contribution because radiative losses, particularly bremsstrahlung, are strongly suppressed for the heavier muon. These two differences largely compensate each other, leaving only the small difference in their total stopping powers. Consequently, both particles deposit nearly the same energy while traversing the silicon and produce closely overlapping partial-absorption responses. The remaining separation between their peak positions arises from the residual stopping-power difference together with particle-transport effects such as multiple scattering, path-length variations, energy-loss fluctuations, and bremsstrahlung-photon escape.

The branching ratio extracted from the singles fit is $\mathrm{BR}(18\;\mathrm{keV}) = (0.033 \pm 0.002\,\mathrm{(stat.)}\,^{+0.003}_{-0.005}\,\mathrm{(syst.)})$ with likelihood-ratio significance exceeding 15$\sigma$ for the most conservative background normalisation. The proton-yield-weighted singles average over the 10--20 keV beam-energy range is $\langle\mathrm{BR}\rangle=(0.0352\pm0.001\mathrm{(stat.)}^{+0.005}_{-0.007}\,\mathrm{(syst.)})$. Both values are consistent with the strongly energy-dependent IPC channel predicted by the near-threshold $0^+$ resonance interpretation of our previous studies\cite{czerski2022prc,czerski2024prc, prx2025}. The 2D coincidence analysis corroborates the singles result within combined uncertainties, with the residual deviations dominated by the model dependence of the threshold-survival correction required for gates lying below the elevated $\Delta E$ threshold.

For the future works, a reduction of the accelerator-chamber electronic-noise level, through improved chamber grounding or mechanical vibration isolation would lower the $\Delta E$ threshold to the level used in the preparation room and unlock the full 2D coincidence acceptance. This would enable a model-independent branching-ratio determination directly from the coincidence dataset, removing the threshold-survival systematic that currently limits the 2D analysis. The response functions and configuration recommendations established in this work are directly applicable to other weak-channel searches in surface laboratories where a continuum-like signal overlaps with cosmic-ray and terrestrial-gamma backgrounds, including low-energy nuclear astrophysics measurements, silicon-substrate background studies for quantum-circuit applications, and low-background charged-particle spectrometry. The same approach, in which run-to-run background variability is converted into a bounded systematic envelope, may also be useful for any precision counting experiment in which the dominant systematic arises from time-varying detector backgrounds.


\section{Acknowledgement}

This work was funded by the National Science Centre, Poland, under the PRELUDIUM grant No. 2025/57/N/ST2/02384.

\bibliography{references}

\end{document}